\def\ls{\lower4pt\hbox{${\buildrel < \over \sim}$}}
\def\gs{\lower4pt\hbox{${\buildrel > \over \sim}$}}

\documentclass[12pt, preprint]{aastex}

\slugcomment{Accepted for publication in {\it The Astrophysical Journal}}

\shorttitle{Modeling BL~Lacertae in 2000}
\shortauthors{B\"ottcher \& Reimer}

\begin{document}

\title{Modeling the Multiwavelength Spectra and Variability of BL~Lacertae in 2000}

\author{M. B\"ottcher\altaffilmark{1}, and A. Reimer\altaffilmark{2}}

\altaffiltext{1}{Department of Physics and Astronomy, Clippinger 339, Ohio University, \\
Athens, OH 45701, USA}
\altaffiltext{2}{Institut f\"ur Theoretische Physik, Lehrstuhl IV, \\
Ruhr-Universit\"at Bochum, D-44780 Bochum, Germany}

\begin{abstract}
BL~Lacertae was the target of an extensive multiwavelength
monitoring campaign in the second half of 2000. The campaign
had revealed optical and X-ray intraday variability on time 
scales of $\sim 1.5$~hours and evidence for significant 
spectral variability both at optical and X-ray frequencies. 
During the campaign, BL~Lacertae was observed in two
different activity states: A quiescent state with relatively
low levels of optical and X-ray fluxes and a synchrotron cutoff
at energies below the X-ray regime, and a flaring state with
high levels of optical and X-ray emission and a synchrotron 
cutoff around or even beyond $\sim 10$~keV. In this paper, 
we are using both leptonic and hadronic jet models to fit 
the broadband spectra and spectral variability patterns 
observed in both activity states in 2000. We start out with 
global spectral models of both activity states. Subsequently, 
we investigate various flaring scenarios for comparison
with the observed short-term variability of BL Lacertae
in 2000. For our leptonic jet model, we find that the
short-term variability, in particular the optical and
X-ray spectral variability, can be best represented with 
a flaring scenario dominated by a spectral-index change of
the spectrum of ultrarelativistic electrons injected into 
the jet. Based on this result, a detailed model simulation
of such a flaring scenario, reproducing the observed optical
and X-ray spectral variability and broadband SED of BL Lacertae
during the BeppoSAX pointing around Nov. 1, 2000, simultaneously,
is presented. Our leptonic modeling results are compared to
fits using the hadronic synchrotron-proton blazar (SPB) model. 
That model can reproduce the observed SEDs of BL~Lacertae in a
scenario with $\mu$-synchrotron dominated high-energy emission.
It requires a significantly higher magnetic field than the
leptonic model ($\sim 40$~G vs. $\sim 2$~G in the leptonic
model) and a lower Doppler factor associated with the bulk
motion of the emission region ($D \sim 8$ vs. $D \sim 18$ in 
the leptonic model). The hadronic model predicts a significantly
larger $\gtrsim 100$~GeV flux than the leptonic models, well
within the anticipated capabilities of VERITAS and MAGIC.

\end{abstract}


\section{\label{intro}Introduction}

BL Lacertae (= 1ES~2200+420; z = 0.069) was historically the prototype 
of the BL Lac class of active galactic nuclei (AGN). These objects are 
characterized by continuum properties similar to those of flat-spectrum 
radio quasars (non-thermal optical continuum, high degree of linear
polarization, rapid variability at all wavelengths, radio jets
with individual components often exhibiting apparent superluminal
motion), but do usually show only weak emission lines (with 
equivalent width in the rest-frame of the host galaxy of 
$< 5$~\AA), if any. In BL~Lacertae itself, however, H$\alpha$ 
(and H$\beta$) emission lines with equivalent widths in excess
of 5~\AA\ have been detected during a period of several weeks in 1995 
\citep{ver95,cor96}, and in 1997 \citep{cor00}. Superluminal motion 
of $\beta_{\rm app}$ up to $7.1 \pm 0.3$ has been observed in this 
object \citep{denn00}.

BL~Lacertae is classified as a low-frequency peaked BL~Lac object 
\citep[LBL;][]{fossati98}. From an interpolation between the GHz 
radio spectrum and the IR -- optical spectrum, it can be inferred 
that its low-frequency spectral component typically peaks at mm 
to $\mu$m wavelengths, while the high-frequency component seems 
to peak in the multi-MeV -- GeV energy range. BL~Lacertae has been 
studied in detail during various intensive multiwavelength campaigns 
\citep[e.g.][]{bloom97,sambruna99,madejski99,ravasio02,villata02,boettcher03}. 
It is a particularly interesting object for detailed X-ray studies since it
is in the X-ray regime where the two broad components of the multiwavelength
SEDs of BL~Lacertae (and other LBLs) are overlapping and intersecting. X-ray 
observations of this source at different epochs show significant flux and 
spectral variability, indicating that the X-ray emission is at times dominated 
by the high-energy end of the synchrotron emission, while at other occasions 
it is dominated by the low-frequency portion of the high-energy bump of the 
SED. In fact, BL~Lacertae has repeatedly shown a concave shape 
\citep[e.g.,][]{madejski99,ravasio02}, with rapid variability mainly 
restricted to the low-energy excess portion of the spectrum 
\citep[e.g.,][]{ravasio02,ravasio03}.

In the framework of relativistic jet models, the low-frequency (radio
-- optical/UV) emission from blazars is interpreted as synchrotron
emission from nonthermal electrons in a relativistic jet. The
high-frequency (X-ray -- $\gamma$-ray) emission could either be
produced via Compton upscattering of low frequency radiation by the
same electrons responsible for the synchrotron emission \citep[leptonic
jet models; for a recent review see, e.g.,][]{boettcher02}, or 
due to hadronic processes initiated by relativistic protons 
co-accelerated with the electrons \citep[hadronic models, for 
a recent discussion see, e.g.,][]{muecke01,muecke03}. The lack
of knowledge of the primary jet launching mechanism and the difficulty
in constraining the jet composition from general energetics considerations
\citep[for recent discussions see, e.g.,][]{sm00,gc01} are currently 
leaving both leptonic and hadronic models open as viable possibilities. 
Also, detailed simulations of particle acceleration at relativistic
shocks or shear layers, which may be relevant for the acceleration 
of ultrarelativistic particles in blazar jets, show that a wide 
variety of particle injection spectra may result in such scenarios 
\citep[e.g.,][]{ob02,so03}, greatly differing from the standard spectral
index of 2.2 -- 2.3, which was previously believed to be a universal
value in relativistic shock acceleration scenarios 
\citep[e.g.,][]{achterberg01,gallant99}. Thus, both the nature of 
the matter in blazar jets and the energy spectra of ultrarelativistic 
particles injected into the emission regions in blazar jets are 
difficult to constrain from first principles. For this reason, we
are leaving these aspects as virtually free parameters in our models,
and attempt to constrain them through the results of detailed 
time-dependent modeling of blazar emission.

While simultaneous broadband spectra are very useful to constrain
both types of blazar jet models, there still remain severe ambiguities 
in their interpretation w.r.t. the dominant electron cooling, injection,
and acceleration mechanisms, as was recently illustrated for the case 
of W~Comae by \cite{bmr02}. Those authors have also demonstrated that 
a combination of broadband spectra with timing and spectral variability 
information, in tandem with time-dependent model simulations 
\citep{krm98,gm98,kataoka00,ktl00,lk00,sikora01,bc02,krawczynski02,kusunose03} 
can help to break some of these degeneracies. For this reason, we had 
organized an intensive multiwavelength campaign to monitor BL~Lacertae 
in the second half of 2000 at radio, optical, X-ray, and very-high-energy
(VHE) $\gamma$-ray frequencies, putting special emphasis on detailed 
variability information. The results of this multiwavelength campaign
were published in \cite{boettcher03}; for more details on the optical
and X-ray observations, see also \cite{villata02} and \cite{ravasio03},
respectively. In \S \ref{observations}, we briefly highlight the main 
results of that campaign, emphasizing those aspects that we will 
specifically use here to constrain our model calculations. 

The purpose of this paper is to use both leptonic and hadronic
jet models to fit the spectral energy distributions (SEDs) and
spectral variability of BL~Lacertae observed in 2000. Following
a brief description of both leptonic and hadronic models in
\S \ref{model_descriptions}, we will first present spectral 
fits to the SEDs of BL~Lac in \S \ref{SED_modeling}. Our 
code used to evaluate leptonic models allows us to make
detailed simulations of the spectral variability resulting from
different plausible flaring scenarios. In \S \ref{var_modeling},
we will first do a general parameter study of various scenarios
and compare the results qualitatively with the observed spectral
variability trends in BL~Lacertae (\S \ref{parameter_study}). 
This will allow us to decide on a preferable flaring scenario, for 
which we subsequently optimize our choice of model parameters to fit 
simultaneously the SED and spectral variability of BL~Lacertae as 
observed in 2000 (\S \ref{var_fit}). Possible physical scenarios
triggering the inferred variability of the electron injection spectrum
will be discussed in \S \ref{discussion}. We summarize in \S \ref{summary}.

Throughout this paper, we refer to $\alpha$ as the energy 
spectral index, $F_{\nu}$~[Jy]~$\propto \nu^{-\alpha}$. A 
cosmology with $\Omega_m = 0.3$ and $\Omega_{\Lambda} = 0.7$ 
and $H_0 = 70$~km~s$^{-1}$~Mpc$^{-1}$ is used.

\section{\label{observations}Summary of observational results}

BL~Lacertae was observed in a co-ordinated multiwavelength 
campaign at radio, optical, X-ray, and VHE $\gamma$-ray energies 
during the period mid-May 2000, until the end of the year. 
Results of the multiwavelength observing campaign have been 
published in \cite{boettcher03}. Here, we briefly highlight the
results that we will specifically concentrate on in our modelling 
effort.

Focusing on an originally planned core campaign period of July 
17 -- Aug. 11, BL~Lacertae was the target of an intensive optical 
campaign by the Whole Earth Blazar Telescope \citep[WEBT;][see also 
{\tt http://www.to.astro.it/blazars/webt/}]{villata00,raiteri01}, 
in which 24 optical telescopes throughout the northern 
hemisphere participated. Details of the data collection,
analysis, cross-calibration of photometry from different 
observatories, etc. pertaining to the WEBT campaign have
been published in \cite{villata02}. BL~Lacertae was in a
rather quiescent state during the core campaign, in which
the densest light curve sampling was obtained. However, the
source underwent a state transition to an extended
high state in mid-September 2000, which lasted throughout 
the rest of the year. 

The WEBT campaign returned optical (R-band) light curves of
unprecedented time coverage and resolution. Brightness variations 
of $\Delta R \sim 0.35$, corresponding to flux variations of 
$(\Delta F)/F \sim 0.4$, within $\sim 1.5$~hr have been found. 
Clear evidence for spectral variability at optical wavelengths
was found, and the color changes were more sensitive to rapid 
variations than the long-term flux level. During well-sampled, 
short flares (on time scales of a few hours), the color changes 
strictly follow the flux variability in the sense that the 
spectra are harder when the flux is higher. A plot of B - R vs. R  
(see Fig. \ref{B_R_fit}) reveals two separate regimes within which 
the R magnitudes are well correlated with the respective B - R 
colors. However, there seems to be a discontinuity at $R \sim 14$~mag, 
separating a high-flux and a low-flux regime. Within each regime, a 
similar range of B - R colors is observed \citep{villata02}.

At X-ray energies, BL~Lacertae was observed with the {\it BeppoSAX}
Narrow Field Instruments (NFI) in two pointings on July 26 -- 27 and 
Oct. 31 -- Nov. 2, 2000 \citep{ravasio03}. In addition, the source 
was monitored by the {\it Rossi X-ray Timing Explorer (RXTE)} Proportional 
Counter Array (PCA) in 3 short pointings per week \citep{marscher04}. 
The details of the {\it BeppoSAX} observations and the data analysis 
methods have been published in \cite{ravasio03}. 

The drastic change of the activity state of BL~Lacertae in 
mid-September observed in the optical range is accompanied by
several large flares in the PCA light curve over a $\sim 2$~months
period, but not by a similarly extended high flux state as seen 
in the optical. In fact, while the average flux level increased 
only slightly, a higher level of activity was indicated by a higher 
degree of variability. 

During our second {\it BeppoSAX} pointing around Nov. 1, 2000,
BL~Lacertae was in an exceptionally bright X-ray outburst state. 
Interestingly, the R-band lightcurve indicates a relatively low 
optical flux, compared to the average flux level after mid-September 
2000, coincident with this X-ray outburst.

In the following, we will concentrate on the data analysis results 
obtained using a neutral hydrogen column density of $N_H = 2.5 \times 
10^{21}$~cm$^{-2}$ and an optical extinction coefficient of $A_B = 
1.42$. During the July 26 -- 27 {\it BeppoSAX} observation, the 
source was in a low flux and activity state. The fit to the 
{\it BeppoSAX}
spectrum resulted in $\alpha = 0.8 \pm 0.1$, 
confirming the low-activity state of the source at that time 
and indicating that the entire X-ray spectrum might have been 
dominated by the low-frequency end of the high-energy component 
of the broadband SED of BL~Lacertae.

The short-term LECS ([0.5 -- 2]~keV) and MECS ([2 -- 10]~keV) 
lightcurves of BL~Lacertae during this observation \citep[see 
Fig. 3 of][]{ravasio03} display a large (factor $> 2$) flare
on a time scale of $\sim 4$~hr, while the source appears less 
variable at higher energies. This behavior has been noted in
this source before \citep[e.g.,][]{ravasio02}, and is even more 
obvious in the Oct 31 -- Nov. 2 observation. 

During the second {\it BeppoSAX} pointing on Oct. 31 -- Nov. 2, 
2000, the LECS + MECS 0.3 -- 10~keV spectrum was well fitted with 
a power-law model with $\alpha = 1.56 \pm 0.03$ \citep{ravasio03}. 
In this observation, BL~Lacertae was also significantly detected 
by the PDS up to $\sim 50$~keV, indicating a spectral
hardening 
in this energy range. The soft shape of the
LECS + MECS spectrum 
clearly indicates that it was dominated by
the high-energy end 
of the low-energy (synchrotron) component in this observation. 
The spectral hardening evident in the PDS
spectrum might indicate 
the onset of the high-energy component
beyond $\sim 10$~keV.

\cite{ravasio03} extracted the LECS and MECS light curves in
three different energy channels during the second {\it BeppoSAX}
pointing, along with the two hardness ratios: HR1 = MECS [2 - 4]
/ LECS [0.5 - 2] and HR2 = MECS [4 - 10] / MECS [2 - 4]. The 
LECS and MECS light curves show significant variability in all
energy channels, with flux variations of factors of $\sim 3$ -- 
4 on time scales down to $\sim 1$ -- 2~hr. 

The X-ray spectral variability on short (intra-day) time scales can
be characterized through variations of the {\it BeppoSAX} hardness 
ratios HR1 and HR2 as a function of the respective count rates. 
An example of such a hardness-intensity diagram (HID) is shown in
Fig. \ref{hid30_fit}. A weak hardness-intensity anti-correlation at 
soft X-rays (HR1 vs. LECS) is generally found. Occasionally, a positive 
hardness-intensity correlation at medium-energy X-rays (HR2 vs. MECS) 
can be found, but is not always apparent.

Generally, no significant cross-correlations with measurable time delays 
between different energy bands could be identified in this campaign. 
A possible correlation between the X-ray and optical light curves with
an optical delay of 4 -- 5~hr during the July 26 -- 27 {\it BeppoSAX}
observation did not hold up to any statistical significance test
\citep[for more details, see][]{boettcher03}. However, if we assume 
that the optical lag of $\sim 4$ -- 5~hr is real and can be interpreted 
as due to synchrotron cooling, it allows an independent magnetic field 
estimate, which is in good agreement with the independent estimate based
on the measured synchrotron peak flux and a basic equipartition argument
(see \S \ref{parameter_estimates}).

\cite{boettcher03} constructed two simultaneous broadband SEDs for the 
times of the two {\it Beppo}SAX pointings (see Fig. \ref{mw_combined}). 
Fig. \ref{mw_combined} illustrates the different activity
states between the July 26/27 and the Oct. 31 -- Nov. 2 {\it Beppo}SAX
observations. In the July 26/27 SED, the synchrotron peak appears
to be located at frequencies clearly below the optical range, and a
synchrotron cutoff near or below the soft X-ray regime. In contrast, the
SED of Oct. 31 -- Nov. 2 shows clear evidence for the presence of the
synchrotron component out to at least 10~keV, and the synchrotron peak
might be located in the optical range. The figure also shows the 
{\it RXTE}~PCA spectrum of the observation a few hours before the 
beginning of the Oct. 31 -- Nov. 2 {\it Beppo}SAX pointing. This 
PCA spectrum shows characteristics rather similar to the low-state 
spectrum, and illustrates the drastic nature of the short-term X-ray 
variability. 

\cite{ravasio03} have shown that the extrapolation of the optical spectrum 
towards higher frequencies does not connect smoothly with the contemporaneous 
soft X-ray spectrum (see their Fig. 5). In addition to the various scenarios
discussed by \cite{ravasio03} to possibly explain this misalignment, it 
seems also possible that it could be an artifact of the flux averaging over 
the $\sim 1.5$~days of the {\it Beppo}SAX observations, including multiple 
short-term flares of only a few hours each. In order to test for this 
possibility, it will be essential to apply fully time-dependent AGN 
emission models, as we do in this paper, and fit time-dependent spectral
variability patterns rather than only time-averaged SEDs.

\section{\label{model_descriptions}Description of leptonic and hadronic
models}

In this section, we give a brief description of the leptonic and hadronic
blazar jet models used for our spectral fitting and variability study
(\S \ref{leptonic_description} and \ref{hadronic_description}). Here, we
also briefly review some general parameter estimates derived previously
\citep{boettcher03} which will be used as a starting point in our spectral 
modelling efforts (\S \ref{parameter_estimates}). 

\subsection{\label{leptonic_description}Leptonic model}

The leptonic model adopted in this paper is a generic, fully time-dependent
one-zone relativistic jet model. The details of the model as well as the
numerical procedure adopted to solve the time-dependent electron continuity
equation and the photon transport equations are given in \cite{bc02}. 
Here, we briefly highlight the salient features of this model.

A population of ultrarelativistic, non-thermal electrons (and positrons) 
is injected into a spherical emitting volume of co-moving radius $R_b$ (the
``blob'') at a time-dependent rate. The injected pair population is specified 
through an injection power $L_{\rm inj} (t)$ and the spectral characteristics 
of the injected non-thermal electron distribution. We assume that electrons 
are injected with a single power-law distribution with low and high energy
cutoffs $\gamma_1$ and $\gamma_2$, respectively, and a spectral index 
$q$.

The jet is powered by accretion of material onto a supermassive central
object, which is accompanied by the formation of an accretion disk with 
a bolometric luminosity $L_D$. The randomly oriented magnetic field $B$ 
in the emission region is parameterized through an equipartition parameter 
$\epsilon_B$, which is the fraction of the magnetic field energy density, 
$u_B$, compared to its value for equipartition with the relativistic 
electron population in the emission region. The blob moves with 
relativistic speed $v / c = \beta_{\Gamma} = \sqrt{1 - 1 / \Gamma^2}$ 
along the jet which is directed at an angle $\theta_{\rm obs}$ (with 
$\mu \equiv \cos\theta_{\rm obs}$) with respect to the line of sight. 
The Doppler boosting of emission from the co-moving to the observer's 
frame is determined by the Doppler factor $D = \left[ \Gamma \, 
(1 - \beta\mu) \right]^{-1}$.

As the emission region moves outward along the jet, particles are 
continuously injected, are cooling, primarily due to radiative losses, 
and may leak out of the system. We parametrize particle escape through 
an energy-independent escape time scale $t_{\rm esc} = \eta \, R_b / c$ 
with $\eta \ge 1$. Radiation mechanisms included in our simulations 
are synchrotron emission, Compton upscattering of synchrotron photons 
\citep[SSC = Synchrotron Self Compton scattering;][]{maraschi92,bm96}, 
and Compton upscattering of external photons (EC = External Compton 
scattering), including photons coming directly from the disk 
\citep{dsm92,ds93} as well as re-processed photons from the broad 
line region \citep{sikora94,bl95,dss97}. The broad line region is 
modelled as a spherical shell between $r_{\rm BLR, in}$ and 
$r_{\rm BLR, out}$, and a radial Thomson depth $\tau_{\rm T, BLR}$. 
$\gamma\gamma$ absorption and the corresponding pair production 
rates are taken into account self-consistently.

\subsection{\label{hadronic_description}Hadronic model}

While leptonic models deal with a relativistic e$^\pm$ plasma in the 
jet, in hadronic models the relativistic jet consists of a relativistic 
proton ($p$) and electron ($e^-$) component. Here we use the hadronic 
Synchrotron-Proton Blazar (SPB-) model of \cite{muecke03} to model the 
spectral energy distribution (SED) of BL Lacaerte in July and November 
2000. 

Like in the leptonic model, the emission region, or ``blob'', in an AGN
jet moves relativistically along the jet axis which is closely aligned 
with our line-of-sight. Relativistic (accelerated) protons, whose particle 
density $n_p$ follows a power law spectrum $\propto \gamma_p^{-q_p}$
in the range $2\leq\gamma_p\leq\gamma_{\rm{p,max}}$, are injected 
instantaneously into a highly magnetized environment ($B =$~const. within 
the emission region), and are subject to energy losses due to proton-photon 
interactions (meson production and Bethe-Heitler pair production), 
synchrotron radiation and adiabatic expansion. The mesons produced in
photonmeson interactions always decay in astrophysical environments.
However, they may suffer synchrotron losses before the decay, which 
is taken into account in this model.

If the relativistic electrons are accelerated together with the protons
at the same site, their injection spectrum shows most likely
the same spectral shape $\propto \gamma_e^{-q_e}$ with $q_e=q_p$.
In the following we shall assume this as a working hypothesis.
The relativistic primary $e^{-}$ radiate synchrotron photons which
constitute the low-energy bump in the blazar SED, and serve as
the target radiation field for proton-photon interactions and
the pair-synchrotron cascade which subsequently develops. The 
SPB-model is designed for objects with a negligible external
target photon component, and hence suitable for BL~Lac objects.
The cascade redistributes the photon power to lower energies where
the photons eventually escape from the emission region. The
cascades can be initiated by photons from $\pi^0$-decay (``$\pi^0$ 
cascade''), electrons from the $\pi^\pm\to \mu^\pm\to e^\pm$ decay 
(``$\pi^\pm$ cascade''), $p$-synchrotron photons (``$p$-synchrotron 
cascade''), charged $\mu$-, $\pi$- and $K$-synchrotron photons 
(``$\mu^\pm$-synchrotron cascade'') and $e^\pm$ from proton-photon 
Bethe-Heitler pair production (``Bethe-Heitler cascade'').

Because ``$\pi^0$ cascades'' and ``$\pi^\pm$ cascades'' generate
rather featureless photon spectra \citep{muecke01,muecke03}, proton 
and muon synchrotron radiation and their reprocessed radiation turn 
out to be mainly responsible for the high energy photon output in 
blazars. The contribution from the Bethe-Heitler cascades is mostly
negligible. The low energy component is dominanted by synchrotron 
radiation from the primary $e^-$, with a small contribution of synchrotron
radiation from secondary electrons (produced by the $p$- and 
$\mu^\pm$-synchrotron cascade). A detailed description of the model 
itself, and its implementation as a (time-independent) Monte-Carlo 
code, has been given in \cite{muecke01} and \cite{reimer04}.

\subsection{\label{parameter_estimates}General parameter estimates}

\cite{boettcher03} have derived some model-independent parameter 
estimates based on the observational results of the BL-Lacertae 
multiwavelength campaign of 2000, which we will briefly summarize 
here. These estimates apply to both leptonic and hadronic models, 
unless specifically noted otherwise.

The co-moving magnetic field can be estimated by assuming that 
the dominant portion of the time-averaged synchrotron spectrum 
is emitted by a quasi-equilibrium power-law spectrum of electrons.
From the observed properties of the synchrotron spectrum,
\cite{boettcher03} have derived a magnetic-field estimate of

\begin{equation}
B_{e_B} = 3.6 \, D_1^{-1} \, e_B^{2/7} \; {\rm G}.
\label{B_eB}
\end{equation}
where $D_1 = D/10$ and $e_B=u_B/u_e$ with $u_e$ the energy density of the relativistic electrons,
and $u_B$ the magnetic field energy density. Typically $e_B\sim 1$ in leptonic models
while $e_B\gg 1$ in hadronic models since $u_p \gg u_e$ ($u_p$ is the
energy density of the relativistic protons) and $u_B \approx u_e + u_p \approx u_p$.
From the modelling in the framework of the SPB model we find $B_{e_B} \approx 28-41$~G
(see Sect. 4.2)
which is in good agreement with the magnetic field values required for this hadronic model.

Although the tentatively identified time delay between the {\it Beppo}SAX 
LECS [0.5 -- 2]~keV and the R-band light curves of $\Delta t^{\rm obs} 
\sim 4$ -- 5~hr was found not to be statistically significant, it is
interesting to investigate which magnetic field could be derived if such
a correlation was indeed real and the delay was caused by synchrotron 
cooling of high-energy electrons. This has been done in \cite{boettcher03},
resulting in

\begin{equation}
B_{\rm delay, RX} = 1.6 \, D_1^{-1/3} \, (1 + k)^{-2/3} \; {\rm G}, 
\label{B_RX}
\end{equation}
where $k = u'_{\rm ext}/u'_{\rm B}$ is the ratio of the external-photon-field
energy density to the magnetic-field energy density in the co-moving frame. 
As pointed out by \cite{boettcher03}, Eq. \ref{B_RX} may, in fact, slightly 
overestimate the actual magnetic field since at least the optical
synchrotron emitting electrons may also be affected by adiabatic
losses and escape. Depending on the details (geometry and mechanism)
of the jet collimation, those processes can act on time scales as
short as the dynamical time scale, which is constrained by the 
observed minimum variability time scale of $\Delta t_{\rm dyn} 
\lesssim 1.5$~hr (in the observer's frame). Another note of caution
that needs to be kept in mind is that the rather large sampling 
time scale of the X-ray light curve of $\Delta t = 1$~hr, precludes
the estimation of magnetic fields larger than $B_{\rm delay, max}
\sim 4.8 \, D_1^{-1/3} \, (1 + k)^{-2/3}$~G from delays between 
the optical and X-ray light curves.

It is remarkable that the two magnetic-field estimates are in good
agreement if the Doppler factor is slightly larger than 10 and/or the 
parameter $e_B$ is only slightly less than 1. We thus
conclude that a magnetic field of $B \sim 2 \, e_B^{2/7}$~G might be a
realistic value for BL~Lacertae. This is also in good agreement with
magnetic-field estimates for this source based on earlier observations
\citep[e.g.,][]{madejski99,ravasio02}.

Based on the magnetic-field estimate of 1.5 -- 2~G for leptonic models, 
the approximate location of the synchrotron peak of the SEDs of BL~Lacertae 
at $\nu_{\rm sy} \sim 10^{14}$~Hz allows us to estimate that the electron 
energy distribution in the synchrotron emitting region should peak at 
$\langle\gamma\rangle \sim 1.4 \times 10^3 \, D_1^{-1/2}$, also in reasonable 
agreement with earlier estimates for this source \citep{madejski99,ravasio02}. 
The location of the synchrotron cutoff in the quiescent state at 
$\nu_{\rm sy, co}^{\rm qu} \lesssim 10^{17}$~Hz then yields a maximum 
electron energy in the quiescent state of $\gamma_2^{\rm qu} \lesssim 
4 \times 10^4 \, D_1^{-1/2}$, while the synchrotron cutoff in the flaring 
state at $\nu_{\rm sy, co}^{\rm fl} \sim 2.4 \times 10^{18}$~Hz yields 
$\gamma_2^{\rm fl} \sim 2 \times 10^5 \, D_1^{-1/2}$. For hadronic models,
the magnetic-field estimate is a factor of $\sim 20$ higher than in the
leptonic case. Consequently, the estimates for the co-moving energies of 
the synchrotron-emitting electrons will be lower by a factor of $\sim 
\sqrt{20} \approx 4.5$.

The superluminal-motion measurements place a lower limit on the bulk
Lorentz factor $\Gamma \gtrsim 8$, and we expect that the Doppler
boosting factor $D$ is of the same order. Since, unfortunately, we
only have an upper limit on the VHE $\gamma$-ray flux during the campaign
of 2000, and no measurements in the MeV --- GeV regime, no independent
estimate from $\gamma\gamma$ opacity constraints can be derived. However,
such an estimate was possible for the July 1997 $\gamma$-ray outburst
and yielded $D \gtrsim 1.4$ \citep{bb00}, which is a much weaker 
constraint than derived from the superluminal motion observations. 

From the optical and X-ray variability time scale, we find an upper
limit on the source size of $R_B \lesssim 1.6 \times 10^{15} \, D_1$~cm.
If the electrons in the jet are efficiently emitting most of their co-moving
kinetic energy before escaping the emission region (fast cooling regime),
then the kinetic luminosity of the leptonic component of the jet would have
to be $L_j^e \gtrsim 4 \pi \, d_L^2 \, (\nu F_{\nu})^{\rm pk} / D^4 \sim
10^{41} \, D_1^{-4}$~ergs~s$^{-1}$. If the electrons are in the slow-cooling
regime (i.e. they maintain a substantial fraction of their energy before
escaping the emitting region) and/or the jet has a substantial baryon
load \citep[see, e.g.,][]{sm00}, the kinetic luminosity of the jet would
have to be accordingly larger. Also, if the magnetic field required to
reproduce the synchrotron emission is present continuously throughout
the jet, the jet luminosity may actually be dominated by the Poynting
flux (see caption of Tab.~\ref{equilibrium_fit}).

In order to estimate the energy density in the external photon field,
an estimate of the average distance of the BLR from the central engine
is required. This can be achieved in the following way. The most recent
determination of the mass of the central black hole in BL~Lacertae can
be found in \cite{wu02}. They find a value of $M_{\rm BH} = 1.7 \times
10^8 \, M_{\odot}$. Then, if the width of the emission lines measured
by \cite{ver95}, \cite{cor96}, and \cite{cor00} is interpreted as due
to Keplerian motion of the BLR material around the central black hole,
we find an estimate of the average distance of the line producing
material of ${\bar r}_{\rm BLR} \sim 4.7 \times 10^{-2}$~pc. With
this value, we can estimate the co-moving energy density in the external
radiation field from the BLR compared to the magnetic-field energy
density as

\begin{equation}
k \equiv {u'_{\rm BLR} \over u'_{\rm sy}} \approx {2 \, L_D \, \Gamma^2
\, \tau_{\rm T, BLR} \over {{\bar r}_{\rm BLR}}^2 \, c \, B^2}
\sim 0.3 {L_{D, 45} \, \Gamma_1^2 \, \tau_{\rm T, BLR, -3} \over
B_{\rm G}^2},
\label{k_estimate}
\end{equation}
where $L_{\rm D} = 10^{45} \, L_{\rm D, 45}$~ergs~s$^{-1}$ is the
bolometric luminosity of the accretion disk, $\Gamma = 10 \, \Gamma_1$
is the bulk Lorentz factor, and $\tau_{\rm T, BLR}$ is the reprocessing
depth of the broad line region. The luminosity of the accretion disk
is very hard to constrain since it has never been observed directly 
in BL~Lacertae. Here, we use a standard value of $L_{\rm D} = 
10^{45}$~erg~s$^{-1}$ as a typical value for moderately luminous
AGN. Using a value of the luminosity of the reprocessed emission 
from the BLR of $L_{\rm BLR} = 4 \times 10^{42}$~ergs~s$^{-1}$ 
\citep{madejski99}, this would imply a value of $\tau_{\rm T, BLR}
\sim 4 \times 10^{-3}$. Then, with a magnetic field of $B \sim 2$~G
in the case of leptonic models,
we find $k \sim 0.3$. Unfortunately, 
the lack of a simultaneous MeV -- GeV $\gamma$-ray observation with 
our data set prevents us
from imposing a tighter constraint on the 
BLR parameters. However,
we point out that our basic model assumptions 
will not be severely
affected by moderate variations in the parameters 
determining $k$. For the case of the leptonic models, \cite{bc02} have 
demonstrated that the spectral and variability patterns observed at 
optical and X-ray frequencies are only very weakly dependent on an 
additional contribution from external Compton scattering, as long as 
that contribution is not strongly dominant over other electron cooling 
mechanisms. In the case of hadronic models, with magnetic fields of 
order $B \sim 30$ -- 40~G (see \S~\ref{hadr_spectra}) the estimate 
on $k$ is lower by more than two orders of magnitude. Under
these 
circumstances, the a priori assumption of a negligible external 
photon field in the hadronic model used here is clearly justified.
Also, because $u'_{\rm sy} \ll u'_{\rm B}$ in the latter models
(see \S \ref{hadr_spectra}) the SSC component does not noticeably 
contribute to the total flux.

The EGRET data from the $\gamma$-ray outburst in 1997, the
highest $\gamma$-ray flux ever observed from this source, are included
in our figures only as a guideline for an upper limit. Those measurements
had been accompanied by simultaneous optical and X-ray observations
\citep{sambruna99,madejski99,bb00}, which indicate that the source
was in a markedly different activity state than during the 2000 campaign.
In particular, the ASCA 2 -- 10~keV X-ray spectrum showed an energy index
of $\alpha = 0.44$, indicating that it might have been entirely dominated 
by the low-energy portion of the high-energy (X-ray -- $\gamma$-ray) 
spectral component. For this reason, we did not make any attempt to 
reproduce the 1997 EGRET data in our model fits.

\section{\label{SED_modeling}Spectral modeling}

\subsection{\label{lept_spectra}Leptonic model fits to the SEDs}

Starting with the parameters derived in \S \ref{parameter_estimates},
we have done a series of simulations with our leptonic jet code, letting
the electron and photon spectra relax to an equilibrium state. Since a
moderate contribution from an external radiation field does not severely
affect the SED and spectral variability signatures at optical and X-ray
frequencies \citep{bc02} and we do not have a measurement of the
MeV -- GeV flux simultaneous with our 2000 campaign data, we set the
BLR Thomson depth to 0 in order to save CPU time. The solid curves in
Fig. \ref{mw_combined} shows our best-fit leptonic models for the two
simultaneous SEDs of July 26/27, and Nov. 1, respectively. The relevant
fit parameters are listed in Tab. \ref{equilibrium_fit}.

The major change of parameters between the quiescent and the flaring
state is given by a hardening of the electron spectrum, both through
a significant change of the injection spectral index $q$ and the 
high-energy cutoff $\gamma_2$. In addition, slight changes in the
Doppler boosting factor $D$ and the injection luminosity $L_{\rm jet}$
are required. 

Tab.~\ref{VHE_fluxes} lists the predicted GeV -- TeV fluxes from our
spectral fits for threshold photon energies $E > 5$~GeV, $E > 40$~GeV,
and $E > 100$~GeV, which have been corrected for $\gamma\gamma$ 
absorption by the intergalactic infrared background absorption
using the models of \cite{aharonian01}. Since we have neglected 
any contribution from external Compton scattering of BLR photons, 
the values listed in Tab. \ref{VHE_fluxes} should be regarded as 
lower limits. The predicted flux levels indicate that BL~Lacertae
should be detectable with the new generation of atmospheric \v Cerenkov
telescope arrays like VERITAS only in an extreme flaring state. If
MAGIC reaches its design goals, it should be able to detect BL~Lacertae
in any state of activity.

We note that in all our leptonic and hadronic fits (see next section),
our model radio fluxes are far lower than the actual data. This is because
our models only follow the evolution of the jet during the early phase of
$\gamma$-ray production during which radiative cooling is strongly dominant
over adiabatic cooling. In this phase, the emission region is highly optically
thick out to GHz radio frequencies. We do not follow the further evolution
of the jet components through a possible phase of expansion in which they
are expected to become gradually optically thin at radio frequencies, because
this would necessitate the introduction of several additional, poorly constrained
parameters. The evolutionary phase of the emission components followed in our
model simulations happens on sub-pc scales, which are not resolveable even 
with VLBI (see, e.g., \cite{denn00} for a recent, detailed discussion of VLBI 
polarimetry of BL~Lacertae) since an angular resolution of 1 mas corresponds 
to a linear scale of $\sim 1.3$~pc at the distance of BL~Lacertae. For this 
reason, our results are consistent with BL~Lacertae being a core-dominated 
radio source even in VLBI images.

\subsection{\label{hadr_spectra}Hadronic model fits to the SEDs}

\subsubsection{\label{hadron_november}Oct. 31 -- Nov. 2}

\begin{figure}
\plotone{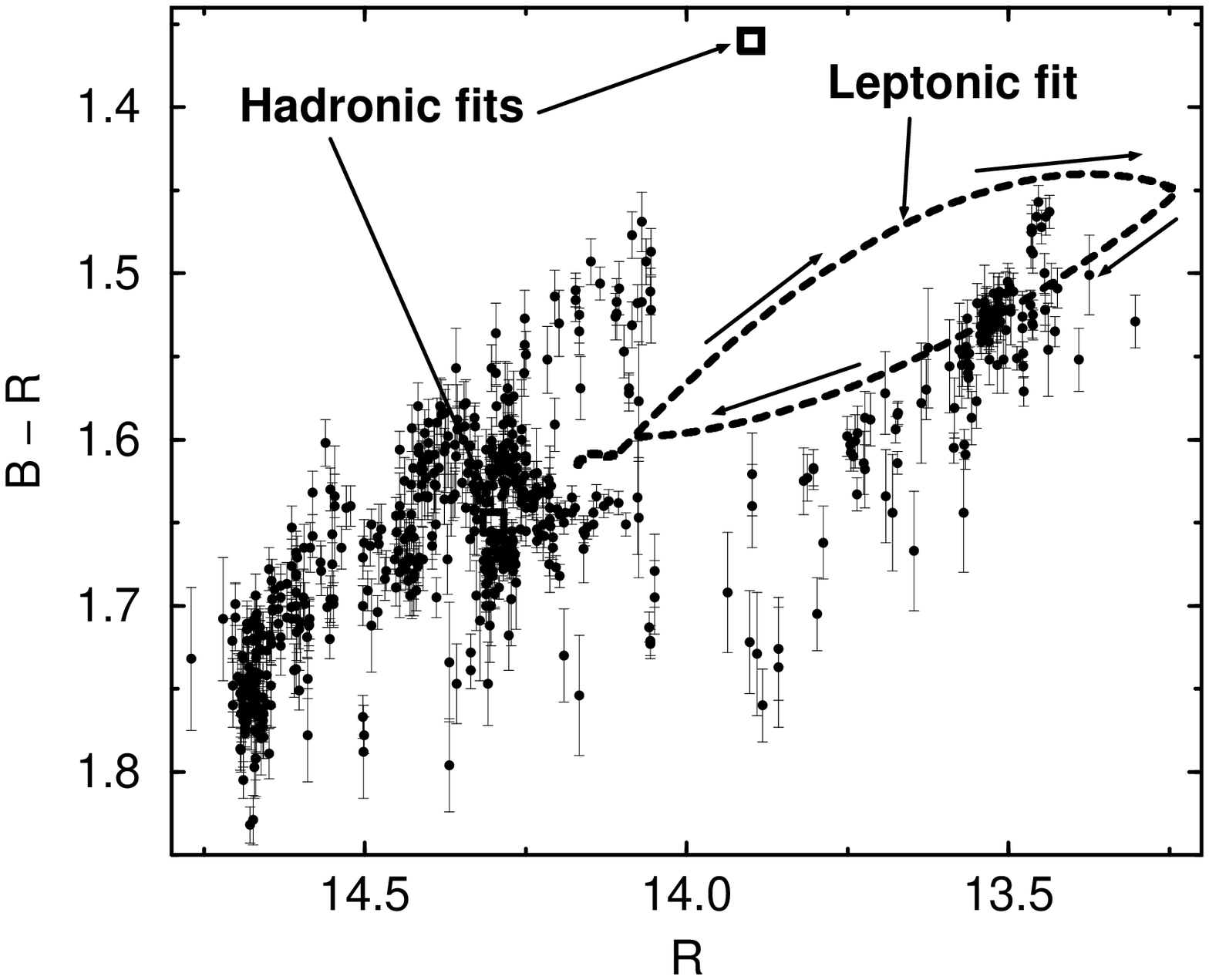}
\caption{Optical B - R color vs. R magnitude of BL~Lacertae in 2000 
\citep[data from][]{villata02}, compared to the result of our best-fit 
model simulation (Fig.~\ref{mw_fit}) with the time-dependent leptonic 
model (dashed curve) and the time averaged emission from our hadronic 
fits (open squares).}
\label{B_R_fit}
\end{figure}

\begin{figure}
\plotone{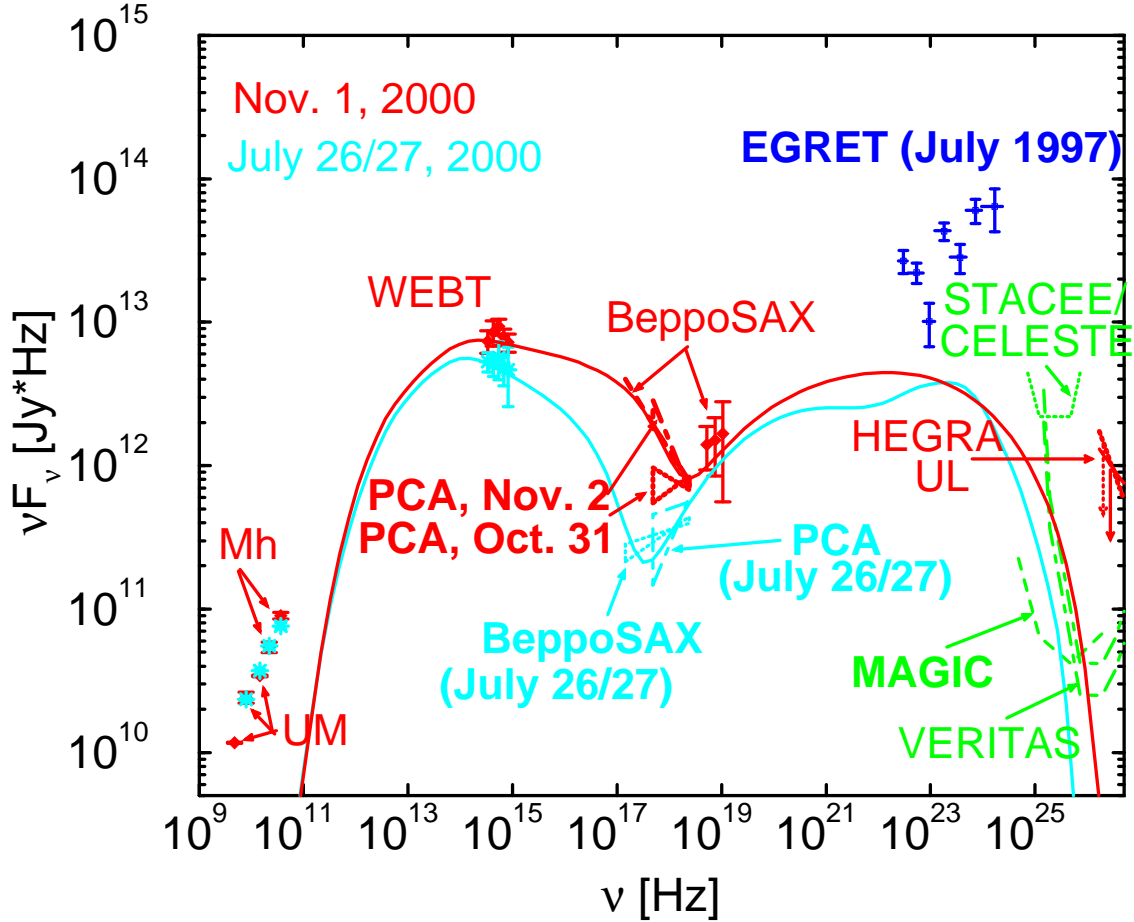}
\caption{Spectral energy distributions of BL~Lacertae on July 26/27, 2000
(stars; cyan in the on-line version; light grey in print), and Oct. 31 -- 
Nov. 2, 2000 (diamonds; red in the on-line version; dark grey in print);
from \cite{boettcher03}. The solid curves show the spectral fits using 
equilibrium solutions of our leptonic synchrotron + Compton model. }
\label{mw_combined}
\end{figure}

\begin{figure}
\plotone{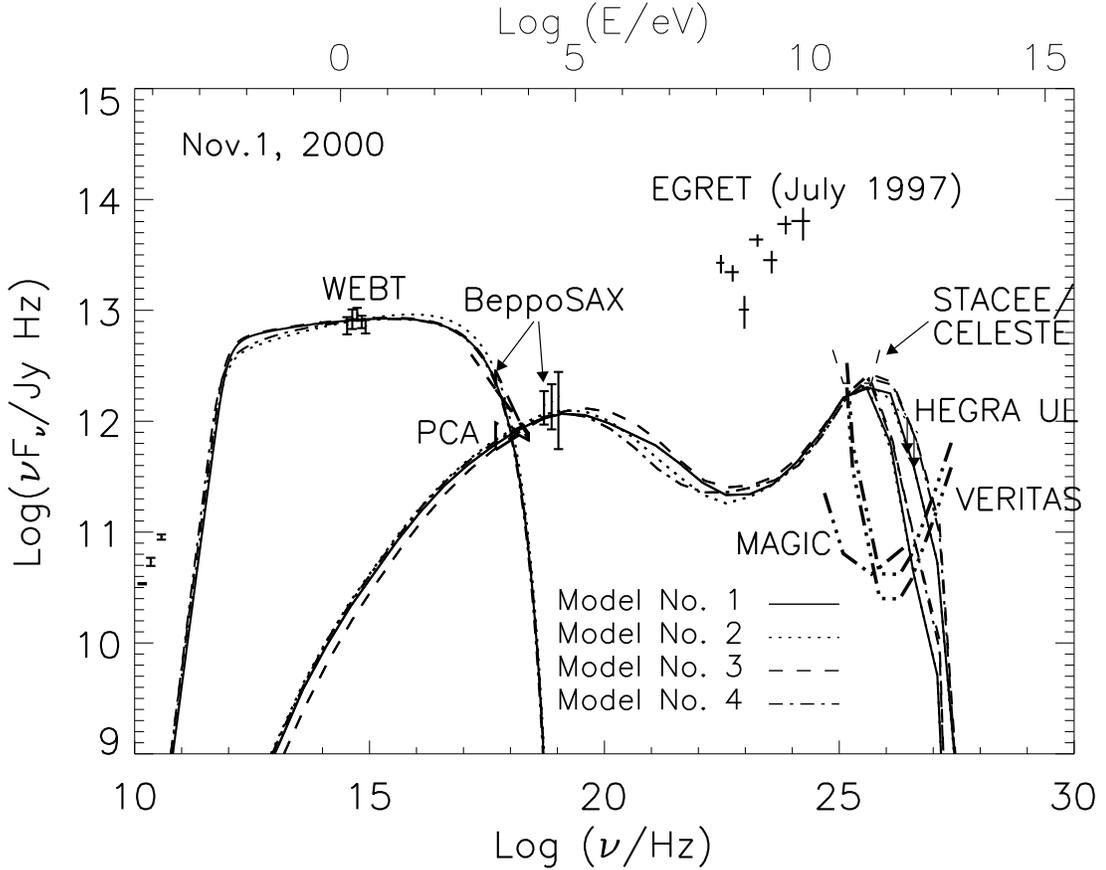}
\caption{Various model fits to the SED of BL Lacertae on November 1,
2000, using the hadronic SPB model. All data and sensitivity limits are corrected
for absorption in the cosmic background radiation field using the background models
of \cite{aharonian01}. The two high frequency branches of the model curves indicate
the resulting fluxes using the two extreme background models of \cite{aharonian01}.
The target photon field for $p-\gamma$ interactions and the pair cascades is the primary electron
synchrotron photon field (solid line at the left). Model parameters are:
$B'=20-40$ G, $D = 9-10$, $R'=
1.5-1.6$$\times$$10^{15}$ cm, $u'_{\rm{phot}} = 5-9$$\times$$10^{11}$ eV
cm$^{-3}$, $u'_p = 36-60$ erg cm$^{-3}$, e/p$\approx$ 1.2-3.2,
$\alpha_e=\alpha_p=1.8-1.9$, L$_{\rm{jet}} \approx 5-8\times
10^{44}$erg/s, $\gamma'_{p\rm{,max}} \approx 1.0-1.5 $$\times$$ 10^{10}$,
$\gamma'_{e\rm{,max}} \approx 2-3 $$\times$$ 10^{4}$.}
\label{SPBfitsNovAll}
\end{figure}

Fig.~\ref{SPBfitsNovAll} shows a summary of SPB-models representing the
data of October 31 -- November 2, 2000, best. The primary electron 
synchrotron spectrum shows a low-energy break at the synchrotron 
self-absorption turnover energy of $\sim \mbox{a few} \times 10^{-3}$~eV,
followed by the synchrotron radiation from the injection particle 
spectrum that is modified by synchrotron losses. The turnover at about 
a few 100~eV with a subsequent steep tail is due to the cutoff in the 
electron distribution at particle Lorentz factor $\gamma_e \sim 10^4$.
This interpretation implies spectral breaks at a few 100~eV energies that 
are larger than 0.5. A spectral break between the optical and X-ray band 
can in principle explain the finding of the optical flux lying significantly 
below the power-law extrapolation of the {\it BeppoSAX} LECS+MECS spectrum 
\citep{ravasio03,boettcher03}. The observed color-flux diagram in the R-band 
(Fig.~\ref{B_R_fit}) constrains the electron injection spectra 
to be not significantly harder than $q_e=1.8$. Our model fits 
(Fig.~\ref{SPBfitsNovAll}) use $q_e = 1.8$ -- $1.9$.

A steep spectral decline at soft X-rays is suggested by the {\it BeppoSAX}
LECS+MECS data. A high magnetic field of $\ge 40$~G in the emission
region leads to a dominance of synchrotron losses throughout the emitted
low-energy component (the escape loss dominated energy range lies below
the synchrotron-self absorption turnover frequency). With these magnetic
field strengths the optical synchrotron emission is expected to lag the
soft X-ray emission by $\lesssim 4$~minutes.

The synchrotron radiation serves as the target photon field for photon-proton
interactions and cascading which determines the radiative output at high
energies. The high energy component of the SED is constrained by the RXTE/PCA
data, the {\it BeppoSAX} PDS data and the $3\sigma$ upper limit from HEGRA.
We have also included the EGRET data from the 1997 outburst (the highest EGRET
flux ever measured from this source) as an upper limit in the MeV-GeV regime.
The hardening of the PDS spectrum seems to indicate the onset of the high energy
component just below 10~keV.

In the SPB model the PDS data can in general be explained by either direct
proton synchrotron radiation or a strong reprocessed cascade component. The
former possibility, however, requires extremely large Doppler factors and/or
very high magnetic field strengths which would increase the total jet power 
to $L_{\rm jet} \sim 10^{47}$~erg/s. Such high values are unlikely for 
low-luminosity BL Lac objects. In the following we therefore concentrate 
on the second option.

\begin{figure}
\plotone{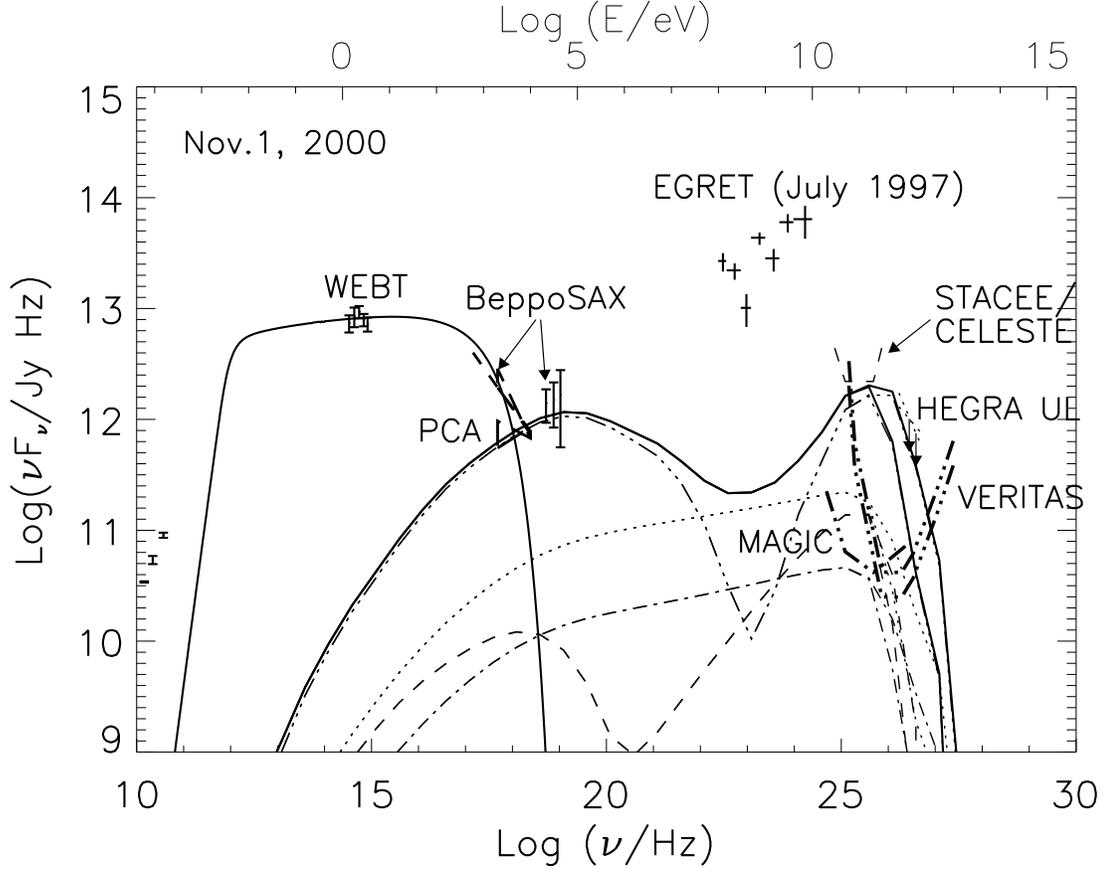}
\caption{Emerging cascade spectra for SPB model 1 from Fig.~\ref{SPBfitsNovAll}.
 The total cascade spectrum (solid line at the right) is the
sum of $p$ synchrotron cascade (dashed line), $\mu$ synchrotron
cascade (dashed-triple dot), $\pi^0$ cascade (dotted line) and
$\pi^{\pm}$-cascade (dashed-dotted line).
All model fluxes are corrected for absorption in the cosmic radiation 
background as described in Fig.~\ref{SPBfitsNovAll}.}
\label{SPBfitsNov}
\end{figure}

No variability has been observed with the PDS within the exposure time of
$\sim 10^5$sec in the jet frame. This constrains the parameter space. If
the hard X-rays are due to reprocessed proton synchrotron radiation, the
magnetic field is limited to values $B \leq 35$~G. For a dominating reprocessed
$\mu$/$\pi$-synchrotron component at hard X-rays only (jet frame) target
photon densities $\geq 10^{10}$~eV~cm$^{-3}$ are in agreement with no
variability within the exposure time. Both requirements favor models with
dominating $\pi$-production loss rates as compared to proton synchrotron
losses. Indeed, all models that fit the November 2000 data exhibit strong 
$\mu$/$\pi$-synchrotron radiation and its reprocessed component while proton 
synchrotron radiation is almost negligible. As an example we show in 
Fig.~\ref{SPBfitsNov} the contributions of the various cascade spectra 
to the total emerging radiation for model 1.

The HEGRA upper limit at $>700$~GeV may potentially limit the maximum proton 
energy. The emerging high energy photon spectrum at source is, however, modified
by photon-photon pair production during propagation through the cosmic background 
radiation field. The optical depth exceeds unity above 0.4 -- 1.2~TeV utilizing
the two most extreme background models in \cite{aharonian01}. This absorption 
effect efficiently prevents photons of energy $>700$~GeV to arrive at Earth.
Another method for estimating the maximum input proton energy is possible through
the luminosity of the reprocessed component provided the target photon density 
allows sufficient reprocessing. In this case, the luminosity of the reprocessed 
component is dependent on the input proton energies. We find that a limit to 
the proton injection spectrum of $\gamma_p < 2\cdot 10^{10}$ (due to
$\pi$-production losses) is in agreement with the observations in the 
X-ray regime.

Reasonable representations of the observed spectral energy distribution (SED)
of November 2000 can be found for Doppler factors $D = 9$ -- 10 (leading to
target photon densities of $5 \ldots 9 \times 10^{11}$~eV~cm$^{-3}$), magnetic
field strengths between 20 and 40~G and electron and proton injection
spectra with spectral indices of $q_e = q_p \approx 1.8$ -- 1.9 (see Table
\ref{equilibrium_fit}). Equipartition is reached within a factor 2. Models
with higher Doppler factors usually violate the upper limit at TeV energies.
In all cases the hard X-ray / soft $\gamma$-ray band up to $\sim 1$~MeV is
dominated by reprocessed $\mu$/$\pi$ synchrotron radiation, which is followed
by a broad ''dip'' up to GeV energies determined by the $\pi^0$-cascade (see 
Fig.~\ref{SPBfitsNov}). GeV -- TeV photons are expected due to $\mu$/$\pi$ 
synchrotron radiation, and may be detectable by $2^{nd}$ generation Atmospheric 
Cherenkov telescopes like VERITAS or MAGIC. Above $\sim 200$~GeV the spectrum 
is noticeably modified by the photons interacting with the cosmic background 
radiation field during propagation. Model fit 3 + 4 are in conflict with the 
HEGRA upper limit at energies $<1$~TeV only for an extremely thin cosmic 
background photon field.

The model parameters representing the data are chosen to satisfy the following
constraints: Flux variability provides an upper limit for the size of the 
emission region. We therefore fix the comoving emission region to $R_b \sim 
c t_{\rm{var}} D \approx 1.6\times 10^{14}$~D~cm for both activity states 
(see Sect.\ref{parameter_estimates}). The range of bulk Doppler factors of 
$D = 7$ -- 10 considered for the fitting procedure is consistent with the 
superluminal motion of $\beta_{\rm app}\approx 7.1\pm 0.3$ detected by 
\cite{denn00}, and imply viewing angles between $\sim 5$ and 8~degrees 
with bulk Lorentz factors $\Gamma=7\ldots 8$. These values are also in 
good agreement with the expectations from unification schemes 
\citep[e.g.,][]{up95}. Furthermore, approximate equipartition between 
particles and fields is anticipated. This effectively constrains the 
magnetic field strengths through Eq.(\ref{B_eB}). The injection spectral 
index $q_e = q_p$ finds limits from the observed optical colors (see 
Fig.~\ref{B_R_fit}). The maximum electron (and proton; see above) 
energy is well constrained by the X-ray observations. In addition, 
the maximum proton energy achievable by acceleration can never exceed 
the limit imposed by the Larmor motion, which must fit into the space 
of the emission region.

\subsubsection{\label{hadron_july}July 26/27}

In comparison to November 2000, BL~Lacertae was in a lower activity state
in July 2000. Fig. \ref{SPBfitsJuly} shows the simultanous broad band data
together with a selection of SPB-models representing this state. Because
both, electrons and protons, are assumed to be accelerated together, the
maximum particle energy of each species reached in this process is expected
to be correlated (though not necessarily linearly). The hard spectrum found
by {\it RXTE} and {\it BeppoSAX} in July 2000 indicates that the radiation
in this band belongs to a separate component from the optical emission
detected by the WEBT campaign, implying a significantly lower cutoff energy 
of the primary electron population in July 2000 than in November 2000
if the magnetic field strength does not change significantly. It follows
that also the maximum proton energies reached in July 2000 should be lower
than in November 2000. Indeed, our modeling procedure requires injected proton
spectra with a high-energy cutoff at lower energies in July 2000 (see Table
\ref{equilibrium_fit}). In addition, we find BL Lacertae's SED in its lower
activity state in agreement with Doppler factors $D = 7$ -- 8. A comparison
with the fit parameters for the November 2000 SED suggests that
the bulk Lorentz factor might be a relevant parameter
for explaing different activity states.

For the modeling of fit 1 -- 4 in July 2000 we use $B = 40$~G, $q_p = q_e =
1.6 \ldots 1.9$, $\gamma_{\rm p, max} = (5 \ldots 9) \times 10^9$ and a primary
electron-to-proton density ratio $n_e / n_p \approx 0.8 \ldots 2.7$.
With Doppler factors $D = 7$ -- 8 the target photon energy density in
the jet frame, $u'_{\rm phot}$, is $\sim (1 \ldots 3) \cdot 10^{12}$~eV~cm$^{-3}$.
The models predict the high energy power output in the GeV-to-TeV regime
due to $\mu^\pm$/$\pi^\pm$-synchrotron radiation altered by $\gamma\gamma$
attenuation in the cosmic backgound radiation field, and a broad ``dip''
in the EGRET energy range determined by the $\pi$-cascades and extending
into the hard X-ray band. The expected flux level at these energies lies
close to EGRET's flux sensitivity (for a typical exposure). The hard X-ray
radiation is due to reprocessed $\mu^\pm$/$\pi^\pm$-synchrotron radiation.
A spectral analysis at 0.01 -- 1~MeV may reveal a broad curvature in the 
spectrum.

While in all these models proton synchrotron radiation plays only a minor role
because of the rather thick target photon field for $p\gamma$-interactions,
we note that also model fits are possible where proton synchrotron emission
is the dominant radiation process from X-rays to GeV $\gamma$-rays. These 
models, however, require large Doppler factors $D \geq 14$ and magnetic 
field strengths $B\sim 60$~G which leads to jet powers that are unreasonably 
high for BL Lac objects.

Models involving meson production inevitably predict neutrino emission due to
the decay of charged mesons. The SPB-model for BL Lacertae in 2000 predict a 
$\nu_\mu+\bar\nu_\mu$-output of about $10^{-8}$~GeV~s$^{-1}$~cm$^{-2}$ peaking
at around $10^{9\ldots 10}$~GeV. The neutrino power at $10^6$~GeV is about
$5\cdot 10^{-12}$~GeV~s$^{-1}$~cm$^{-2}$. Neutrino flavor oscillations
are not included in these estimates.

In summary, the hadronic SPB-model predicts TeV-emission on a flux level near 
or below the detectability capabilities of CELESTE and STACEE for BL~Lacertae,
but clearly above the sensitivity limit of future instruments like VERITAS, 
MAGIC and H.E.S.S. While leptonic models predict integral fluxes at $>5$~GeV
for BL Lacertae on a similar level as hadronic models do,
(sub-)TeV emission detectable with very high-sensitivity instruments is only
predicted for the hadronic emission processes, in contrast to leptonic 
models (see Tab.2). Interestingly, this finding is similar to the case of W Comae
in 1998 where a similar comparative study has been performed \citep{bmr02}.
High-sensitivity TeV observations may therefore be useful as a diagnostic 
to distinguish between the hadronic and leptonic nature of the high-energy 
emission at least from some LBLs, in addition to its possible neutrino 
emission.

\section{\label{var_modeling}Spectral variability in the leptonic model}

Due to the multitude of parameters involved in our models, we may expect 
that our choice of parameters is not unique. The ambiguities in pure
spectral modeling of blazar SEDs have been drastically demonstrated for 
the case of W~Comae by \cite{bmr02}. In order to refine our choice of 
parameters for our leptonic model fit and investigate the source of 
variability of BL~Lacertae, we have done a detailed parameter study 
of various plausible flaring scenarios, starting from parameters of 
our quiescent-state fit. The results of this parameter study have been 
compared qualitatively with the observed trends in BL~Lacertae in 2000 
in order to pin down the most likely flaring scenario at work in this 
source (see \S \ref{parameter_study}). Based on this result, we have 
then resumed our fitting procedure to fit simultaneously the SED and
optical and X-ray spectral variability patterns consistently in one 
complete model (\S \ref{var_fit}).

\subsection{\label{parameter_study}Parameter study on spectral variability}

The variability of blazars can in principle be initiated by a multitude
of physical processes, all of which would imply specific changes in the
fundamental modeling parameters of leptonic jet models. For the purpose
of a qualitative comparison with the observed spectral variability patterns
of BL~Lacertae, we have done a series of simulations, focusing on a fluctuation
of one or 2 of the basic model parameters, leaving all other parameters
unchanged: (a) the total injection luminosity of ultrarelativistic particles
into the jet $L_{\rm jet}$, (b) the injection spectral index $q$, 
(c) the high-energy cutoff of the electron injection spectrum $\gamma_2$,
and (d) a combination of electron spectral hardening from (b) and (c). 
Other scenarios like a fluctuation in the Doppler factor $D$ or the
magnetic field $B$ only can be ruled out immediately by virtue of the 
observed spectral variability. 

In our simulations, we have represented a parameter fluctuation by a
change to a new parameter value over a time $\Delta t'_{\rm flare} = 
2 \, R_B / c$, and then switching back to the equilibrium value. In 
the case of simulations (b) -- (d) there is still an ambiguity concerning 
the choice of the normalization of the electron injection spectrum under 
spectral fluctuations. We have executed the suite of simulations (b) -- 
(d) for two extreme assumptions: (1) leaving the injection power 
$L_{\rm inj}$ constant between the equilibrium state and the simulated 
flare, and (2) leaving the total number of injected electrons per unit 
time constant. We have found that the optical and X-ray spectral variability
patterns for those two cases do not differ substantially from each other. 
The results presented in the following paragraphs concerning fluctuating
electron spectral parameters refer specifically to the case of unchanged 
$L_{\rm inj}$. 

A typical set of simulation results is illustrated in Figs. \ref{sed_34}
and \ref{lc_34}. From the simulated, time-dependent spectra and light 
curves, we have calculated optical R-band magnitudes and color indices
B -- R. We have folded the simulated X-ray fluxes through the detector
response of {\it BeppoSAX}, using the exact same effective area curves 
as used in the data analysis of \cite{ravasio03} to evaluate the 
resulting {\it BeppoSAX} count rates and hardness ratios as mentioned 
in \S \ref{observations}. The simulated optical and X-ray spectral 
variability patterns from our flaring scenarios (a) -- (d) are compared 
in Fig. \ref{sv_comparison}. 

First, we note that a model with a fluctuation of only the injection 
luminosity (a) is predicting very limited X-ray spectral variability 
and does not lead to the characteristic, positive brightness -- hardness 
correlation observed at optical frequencies. Such a scenario thus seems 
unlikely to be the driving mechanism behind the variability of BL~Lacertae. 

Our model simulation with a fluctuation of the electron spectral index
$q$ only (b) qualitatively reproduces the optical color -- magnitude
relation and the hardness -- intensity anti-correlation at soft X-rays.
It appears to be capable of reproducing a weak positive hardness -- 
intensity correlation at harder X-rays (HR2 vs. MECS 4 -- 10~keV count 
rate), which has occasionally been observed during our campaign. 
We conclude that such a scenario has a good potential to reproduce
all the optical and X-ray spectral variability patterns observed during
the 2000 campaign on BL~Lacertae.

A scenario invoking primarily a fluctuation in $\gamma_2$ (c) predicts
a very small amplitude of optical variability, compared to the X-ray
variability amplitude. It does predict a strong flux -- hardness 
anti-correlation at soft X-rays, as observed in BL~Lacertae, but 
fails to reproduce the optical color -- magnitude correlation. We 
therefore conclude that this mechanism is not consistent with the
observed spectral variability of BL~Lacertae either. A scenario of
combined changes of $q$ and $\gamma_2$ (d) does qualitatively reproduce
both the optical color -- magnitude correlation and the soft X-ray
hardness -- intensity anti-correlation, but also predicts a strong
hardness -- intensity anti-correlation at harder X-rays (HR2 vs.
MECS 4 -- 10~keV count rate), which has not been observed by 
{\it BeppoSAX}. 

In summary, we find that our flaring scenario (b), based on a hardening
of the electron injection spectral index $q$ only, seems to be the most
promising candidate for modeling the SED and spectral variability of
BL~Lacertae. 

\subsection{\label{var_fit}Simultaneous SED + variability model}

We are now ready to narrow down the parameter choices to model
simultaneously the SED and spectral variability of BL~Lacertae
in 2000. For this purpose, we are first choosing parameters 
similar to the low state of July 26/27, but with the higher
Doppler factor of $D = 18$ to achieve approximate agreement 
with the average optical flux level around Nov. 1 and the hard 
{\it RXTE} PCA spectrum measured just prior to the flaring
episode caught during the second {\it Beppo}SAX pointing on
Oct. 31 -- Nov. 2, 2000. Various scenarios of short-term 
fluctuations of the electron spectral index over $\Delta t'_{\rm flare}
= 2 \, R_B/c$ were tested and compared with the observed SED,
optical color -- magnitude correlation and the {\it Beppo}SAX
hardness -- intensity correlations for individual flares during
the Oct. 31 -- Nov. 2, 2000, pointing. 

Satisfactory agreement with all three of these observational
results was achieved for the following choice of parameters: 
$D = 18$; $\gamma_1 = 1000$, $\gamma_2 = 5 \cdot 10^4$, $q = 3$ 
outside the flaring episode, changing to $q \to 2.4$ during the 
flare; $L_{\rm jet} = 2.5 \times 10^{40}$~ergs~s$^{-1}$, 
$\epsilon_B = 1$, yielding a magnetic field of $B = 2.0$~G. 
The broadband spectral evolution resulting from this flaring
scenario is illustrated in Fig. \ref{mw_fit}. It indicates
how this flaring scenario reproduces the hard X-ray spectrum
seen by PCA right before the flaring episode, and switches to
the synchrotron-dominated soft high-flux spectrum during the
flare. The light curves at optical, X-ray and $\gamma$-ray 
frequencies resulting from this simulation are shown in 
Fig.~\ref{lightcurve_fit}. The significantly larger flaring
amplitude at X-rays compared to optical frequencies is clearly
well reproduced. The flaring amplitude is largest at the
highest $\gamma$-ray energies, where the flux increases by
almost 2 orders of magnitude, to reach levels well above the
anticipated, nominal detection threshold of MAGIC.

The results of our leptonic fit simulation are compared to the 
observed optical color -- magnitude correlation and to the time
averaged emission from our hadronic fits in Fig.~\ref{B_R_fit}.
We see that the hadronic fit for the low state lies well within
the observed range of optical colors and R-band magnitudes, while
the optical color predicted from the high-state fit is harder by
$\Delta (B - R) \sim 0.25$ (corresponding to a difference in the 
local spectral index of $\Delta\alpha \sim 0.5$) than the observed 
B - R values. The leptonic flaring fit coincides reasonably well 
with the range of R magnitudes and B - R colors observed during 
the active phase after Sept. 2000, though the actual simulated 
spectral hysteresis curve lies slightly above the observed correlation.
We have explored multiple attempts to remedy this slight discrepancy,
but could not find a better representation of the data than the one
shown in Fig.~\ref{B_R_fit} which would still be consistent with both
the SED and the X-ray variability patterns discussed below. However,
the difference is minute --- even for the hadronic model fit ---, 
and may be explained by uncertainties in the adopted de-reddening 
and the subtraction of the host galaxy contribution \citep[for an 
in-depth discussion of these issues, see][]{villata02,ravasio03}.

Fig.~\ref{hid30_fit} compares our simulated X-ray spectral hysteresis
curves to the observed hardness-intensity correlation during a 
well-resolved X-ray flare at 0.5~hr -- 6.5~hr UT on Nov. 1, 2000. 
The figure illustrates that the overall flux levels and hardness
ratio values are well within the observed range, and that the
time evolution of those values is consistent with the results of
our simulation. Clearly, the statistical errors on the {\it Beppo}SAX 
count rate and hardness-ratio measurements are too large to test for
the existence of the actual spectral hysteresis phenomena predicted
in our simulation. Our fit predicts slight counterclockwise spectral
hysteresis for our favoured SED + spectral variability fit. Future
observations using, e.g., {\it Chandra} or {\it XMM-Newton} would be
extremely useful to test this prediction.

Finally, we discuss a possible 4 -- 5~hr delay of the optical 
fluxes with respect to X-ray flares for which a statistically
not significant hint in the BL~Lac data of 2000 was found. All of our
simulations discussed in this section did not lead to a systematic 
time delay with significant flux peak separation in time between the 
X-ray and optical flares. Such a feature might be expected in a scenario 
where solely a high-energy population of electrons is injected into the
jet, which subsequently cools due to radiative losses. We have run 
a simulation, similar to the ones described above, but injecting
only a narrow distribution of ultrarelativistic electrons into the
jet during the flare. Such a scenario does reproduce the spurious 
4 -- 5~hr delay, but would predict a strong anti-correlation between 
optical flux and hardness, in contradiction with the observed 
color-magnitude relation. Thus, we conclude that such a scenario 
can be ruled out.

\section{\label{discussion}Discussion}

Overall, our parameter choice for the leptonic models for BL~Lacertae
are in reasonable agreement with those found by other authors based 
on earlier multiwavelength campaigns on this source 
\citep[e.g.,][]{madejski99,bb00,ravasio03}. In agreement with those 
authors, our best-fit Doppler factor of $D = 18$ is well within the 
range typically found for blazar modelling, and the magnetic field 
of $B = 2$~G is intermediate between typical values found for leptonic 
modeling of flat-spectrum radio quasars (FSRQs) and high-frequency 
peaked BL~Lac objects (HBLs). While FSRQs are usually successfully 
modelled with $B \gtrsim $~a few G 
\citep[e.g.,][]{vmontigny97,sambruna97,mukherjee99,ghisellini99,hartman01}, 
typical values found for HBLs are of the order of $B \lesssim 0.1$~G 
\citep[e.g.][]{tmg98,kataoka99,ca99,kataoka00,petry00,krawczynski02}. 

In our analysis of the spectral variability patterns, we have found that
those patterns can successfully be modelled with a fluctuation of the
electron injection spectral index. Remarkably, our time-dependent fits
indicate that an injection index larger than $q \sim 2.3$, even during 
the peak of an individual short-term flare, is required.
If the injection 
of ultrarelativistic electrons into the emitting volume
is caused by 
Fermi acceleration at relativistic shocks, detailed numerical
studies 
have shown that with fully developed turbulence in the downstream
region, 
a unique asymptotic index of $q \sim 2.2$ -- 2.3 should be expected
\citep[e.g.,][]{achterberg01,gallant99}. However, recently \cite{ob02} have
shown that Fermi acceleration might lead to drastically steeper injection
spectra if the turbulence is not fully developed. Furthermore, depending on
the orientation of the magnetic field at the shock front, an abrupt steepening 
of the injection spectra may result if the shock transits from a subluminal 
to a superluminal configuration. In this context, our leptonic fit results 
may indicate that such predominantly geometric effects, may be the cause
of the rapid variability observed in BL~Lacertae.

In their analysis of the {\it BeppoSAX} + optical continuum spectra of the
Oct. 31 -- Nov. 2 observations, \cite{ravasio03} have noticed that the time
averaged optical and LECS + MECS X-ray spectra can not be connected smoothly
using a single power-law or a smoothly connected broken power-law. They have
suggested and investigated several possibilities how this discrepancy could 
be remedied, including a variable dust-to-gas ratio, the bulk Compton 
process \citep{sikora97}, a multi-component emission model, and a spectral
upturn resulting from Klein-Nishina effects on the electron cooling rates.
The results of our combined leptonic spectral + variability modeling of 
BL~Lacertae suggest that the discrepancy ultimately arises artificially 
as a result of the time averaging involved when producing the high-quality 
{\it Beppo}SAX LECS + MECS spectrum. The fact that our time-dependent 
leptonic fits reproduce the observed ranges of optical and X-ray fluxes 
and spectral indices simultaneously resolves the issue of this optical -- 
X-ray spectral discrepancy, and removes the need for any additional 
assumptions concerning intergalactic absorption and/or electron populations 
in the jet.

In contrast to leptonic models, the hadronic SPB model required significantly 
larger magnetic field strengths (of order 20 -- 40~G) on the length scale 
of the size of the emission region of $\sim 10^{15}$~cm. The range of 
Doppler factors appears about a factor of 2 lower than in the leptonic 
models. The total jet power, which turns out to be below the estimated 
accretion disk luminosity, remains larger in hadronic models than in leptonic 
ones owing to the higher particle and field energy content. The state transition 
from low to higher activity in 2000 is well described by an increase of the 
co-moving particle energy and the bulk Lorentz factor. In the picture of 
diffusive shock acceleration (in the test particle limit) the maximum 
particle energies are related to the magnetic turbulence spectrum 
\citep[see, e.g.,][and references therin]{bs87,reimer04}. The required 
maximum electron and proton energies in the presented models can be 
understood if the particles gain energy by diffusive shock acceleration
in a $\propto k^{-1.1\ldots -1.3}$ turbulence spectrum where $k$ is the 
wave number in the turbulent magnetic field. Spectral-index changes can 
not be ruled out, but they are not the dominant cause of the spectral 
variability of BL~Lacertae in the framework of the presented modeling 
using the SPB model.

An interesting diagnostic for the particle content in the jet -- in addition 
to any neutrino detections -- might be achieved through high sensitivity 
observations in the (sub-)TeV energy range by e.g. MAGIC or VERITAS. While 
both, leptonic and hadronic models, predict a similar flux level in the 
GeV-energy range, hadronic models predict about an order of magnitude higher 
flux values than leptonic ones do above 40 GeV for BL Lacerate in 2000.
Furthermore, considering the results of our variability study, the predicted
VHE $\gamma$-ray flux from a leptonic jet only reaches the peak level mentioned
above during short flares. It depends critically on the duty cycle of such
flaring events whether a sufficient time-averaged level of VHE flux can be
sustained for low-energy-threshold Cherenkov telescopes like MAGIC to 
accumulate a measurable signal.

Our leptonic SED + spectral variability fit predicted spectral hysteresis
at soft X-ray energies which might serve as a confirmation of our fit results. 
The limited count statistics of our {\it Beppo}SAX observations did not 
firmly establish nor rule out the existence of X-ray spectral hysteresis. 
More sensitive, dedicated observations by {\it Chandra} and/or 
{\it XMM-Newton} would be extremely helpful to test this prediction. 
According to our hadronic model fits presented in this paper, flares of 
BL~Lacertae were primarily caused by increasing particle energies. If 
this is indeed the dominant flaring mechanism and it is not accompanied 
by significant changes of the electron spectral index, then short-term X-ray 
spectral variability might be reasonably well represented by the patterns 
resulting from our leptonic models with increasing $\gamma_2$ with at most 
very moderate spectral-index changes. These did not show significant spectral 
hysteresis. However, fluctuations of the electron injection spectral index
could not be excluded in our hadronic fits. Consequently, the presence of 
pronounced soft X-ray spectral hysteresis in BL~Lacertae may slightly favour
the leptonic models, while its absence would favour hadronic models, but
the discriminating power of such a measurement in the case of BL~Lacertae
would be rather limited. 

\section{\label{summary}Summary}

In this paper, we have presented the results of detailed numerical modeling 
of the SEDs and spectral variability of BL Lacertae in 2000, using both 
leptonic and hadronic (SPB) jet models. Details of the data analyses and 
observational results have been published in three previous papers on 
this campaign \citep{villata02,ravasio03,boettcher03}. The main results
of our modelling work are:

\begin{itemize}

\item Both leptonic and hadronic models are able to provide acceptable
fits to the SEDs of BL~Lacertae in 2000, both in the low activity state
on July 26/27 and in the high activity state on Oct. 31 -- Nov. 2.

\item In addition to the naturally much higher overall jet power required
for hadronic models ($\sim 6 \times 10^{44}$~ergs~s$^{-1}$ vs. $\lesssim 6 
\times 10^{42}$~ergs~s$^{-1}$ [depending on the possible Poynting-flux
contribution]), the hadronic SPB model requires a factor of $\sim 20$ higher 
magnetic fields ($\sim 30$ -- 40~G vs. $\sim 2$~G) and a significantly lower 
bulk Lorentz factor ($\sim 7$ -- 9 vs. $\sim 18$). 

\item Considering time-averaged emission during the two intensity states,
hadronic models predict a sustained level of multi-GeV -- TeV emission which 
should be detectable with second-generation atmospheric Cherenkov telescope 
systems like VERITAS, HESS, or MAGIC. In contrast, our leptonic model only 
predicts a peak flux exceeding the anticipated nominal MAGIC sensitivity 
during short flares; the accumulated fluence over observing time scales of 
several hours might not be sufficient for a significant detection. Thus,
a future VHE detection of BL~Lacertae would be a strong indication for 
hadronic processes being at work in this object.

\item A parameter study of various spectral variability scenarios in the
framework of our leptonic jet model revealed that the observed optical and 
X-ray spectral variability in BL~Lacertae in 2000 can be reproduced through
short-term fluctuations of only the electron injection spectral index, with 
all other parameters remaining unchanged. Our simulation of this scenario 
predicts counter-clockwise spectral hysteresis at X-ray energies. Such 
hysteresis was not predicted in the specific SPB model fits presented in
this paper, but could not clearly be ruled out either. Thus, sensitive 
spectral-hysteresis measurements of BL~Lacertae could possibly serve as 
a test of our modeling results and a secondary diagnostic to distinguish 
between leptonic and hadronic models, though, by itself, it would not be
sufficient as a model discriminant. 

\item The previously noted discrepancy between the time-averaged optical 
and X-ray spectra may be resolved by considering the spectral variability. 
Our successful modeling of the observed time-dependent flux and hardness 
values at optical and X-ray frequencies in the framework of a leptonic model
effectively removes the need for additional assumptions concerning additional
particle populations, extreme Klein-Nishina effects on the electron cooling
rates, and/or anomalies in the intergalactic absorption.

\end{itemize}

\acknowledgments
We thank M. Ravasio for sending us the BeppoSAX effective area curves
used in the data analysis of the BL~Lac 2000 campaign, and the anonymous
referee for a constructive report which has definitely helped to improve 
the clarity of the manuscript.
AR thanks the Bundesministerium f\"ur Bildung und Forschung
for financial support through DESY grant Verbundforschung 05CH1PCA6.

\newpage

\newpage


\newpage

\newpage

\newpage

\begin{figure}
\plotone{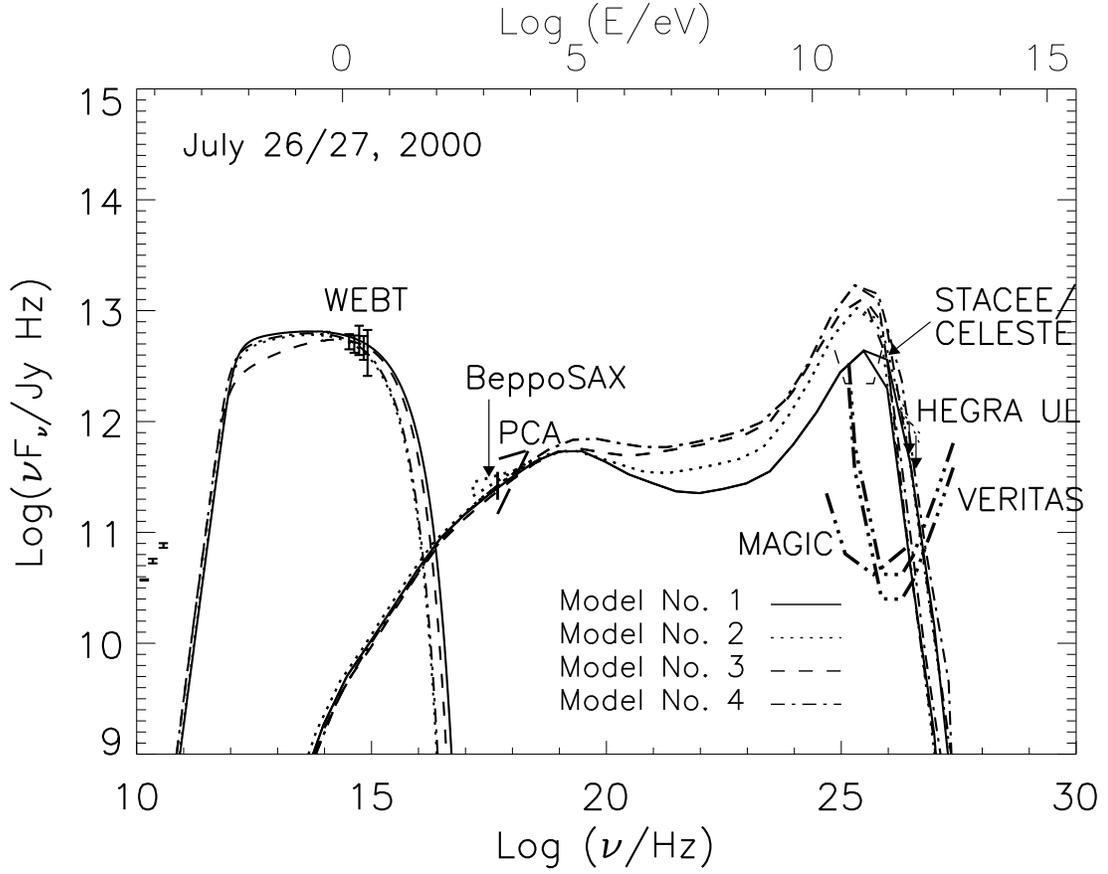}
\caption{Various model fits to the SED of BL Lacertae on July 26/27, 2000,
using the hadronic SPB model. See caption of Fig. \ref{SPBfitsNov} for explanations.
Model parameters are: $B'=40$ G, $D = 7-8$, $R'=
1.1-1.3$$\times$$10^{15}$ cm, $u'_{\rm{phot}} = 1-3$$\times$$10^{12}$ eV
cm$^{-3}$, $u'_p = 270-300$ erg cm$^{-3}$, e/p$\approx$ 0.8-2.7,
$\alpha_e=\alpha_p=1.6-1.9$, L$_{\rm{jet}} \approx 6\times
10^{44}$erg/s, $\gamma'_{p\rm{,max}} \approx 5-9 $$\times$$ 10^{9}$,
$\gamma'_{e\rm{,max}} \approx 1.6-2.4 $$\times$$ 10^{3}$.}
\label{SPBfitsJulyAll}
\end{figure}

\newpage

\begin{figure}
\plotone{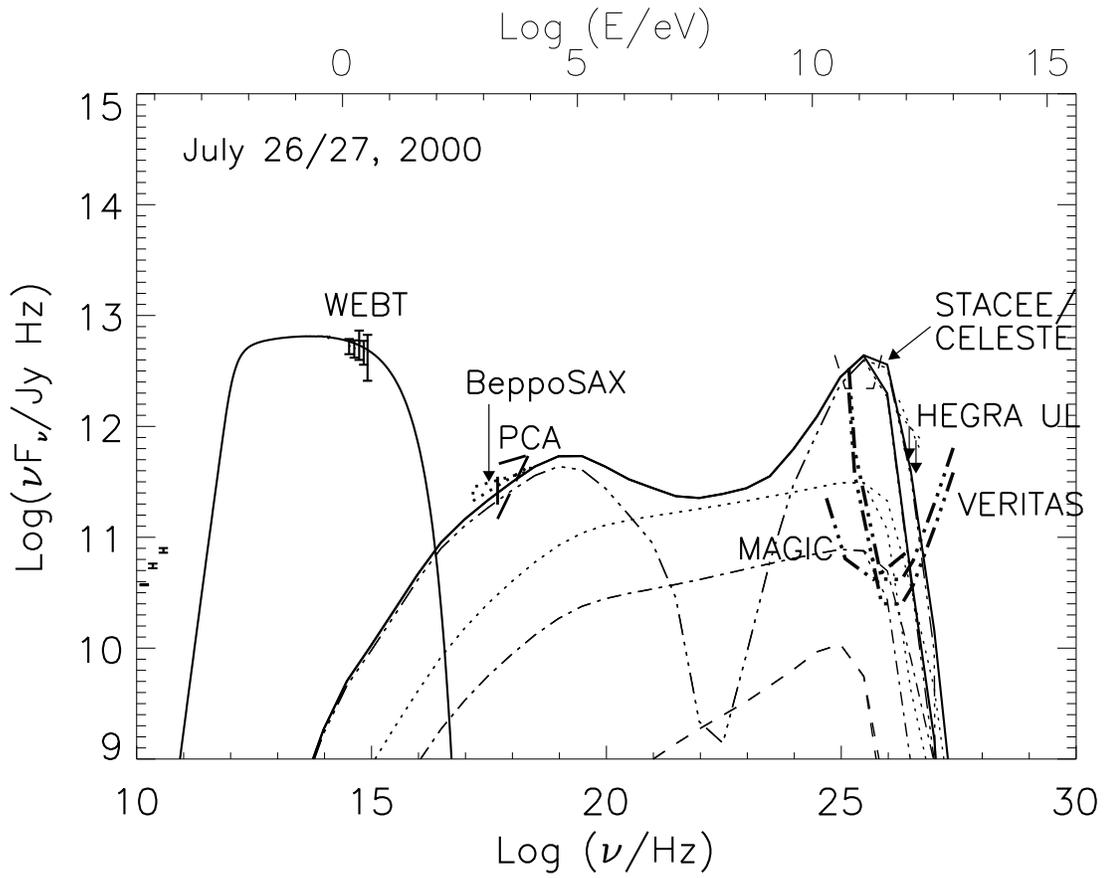}
\caption{Emerging cascade spectra for SPB model 1 from Fig.~\ref{SPBfitsJulyAll}.
See caption of Fig. \ref{SPBfitsJulyAll} for explanations.}
\label{SPBfitsJuly}
\end{figure}

\newpage


\begin{figure}
\plotone{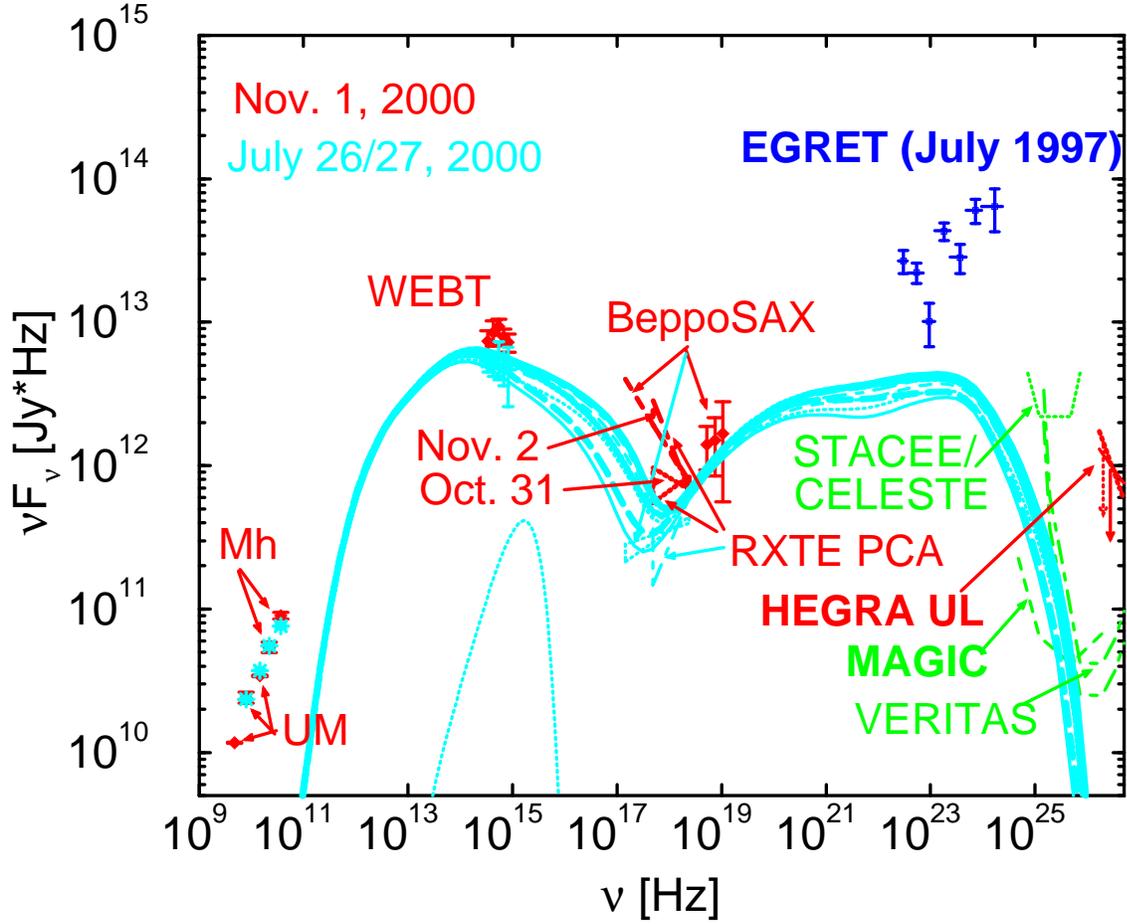}
\caption{Time-dependent model spectra for the case of a fluctuation in
electron injection index $q = 2.5 \to 2.3$ and high-energy cutoff 
$\gamma_2 = 2 \times 10^4 \to 4 \times 10^4$. Other parameters as for
the fit to the quiescent state (see Tab. \ref{equilibrium_fit}). Time
sequence is: thin solid $\to$ thin dotted $\to$ thin long-dashed $\to$
thin dot-dashed $\to$ thin dashed $\to$ thick solid $\to$ thick dotted
$\to$ thick long-dashed; equi-distant time steps of $\Delta t_{\rm obs}
= 1.2$~hr.}
\label{sed_34}
\end{figure}

\newpage

\begin{figure}
\plotone{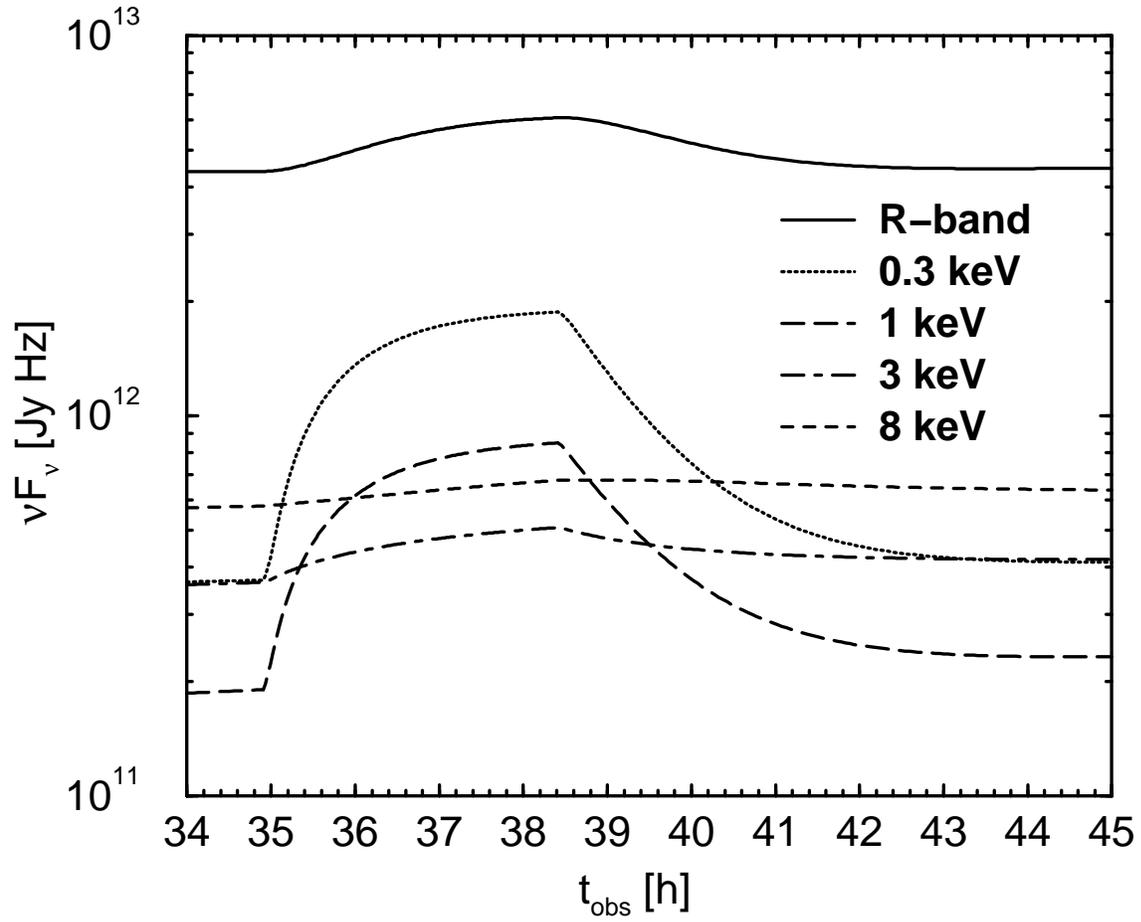}
\caption{Simulated light curves at optical and X-ray frequencies from the
simulation illustrated in Fig. \ref{sed_34}. }
\label{lc_34}
\end{figure}

\newpage

\begin{figure}
\plotone{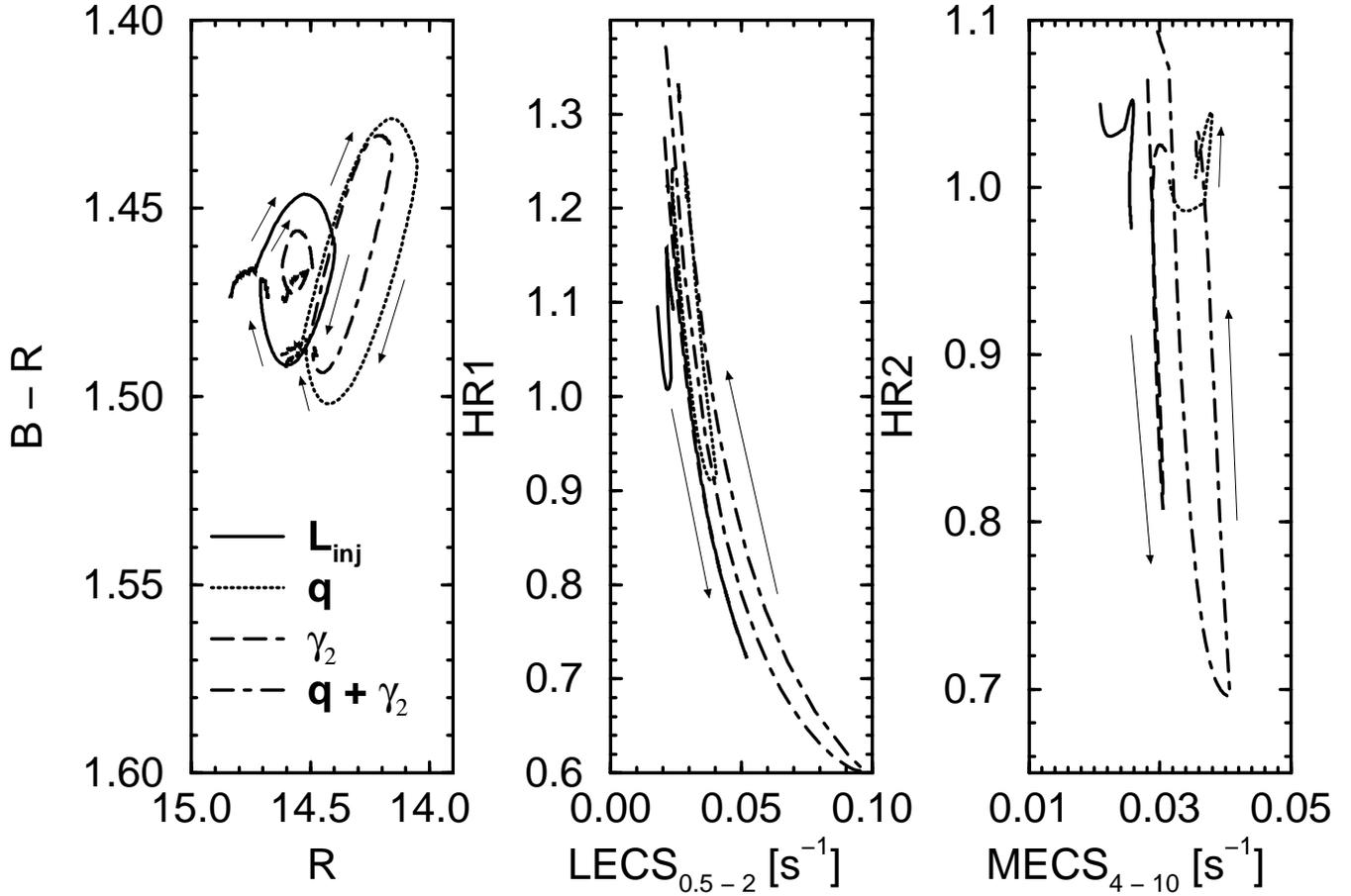}
\caption{Comparison of the optical and X-ray spectral variability patterns 
for various generic flaring scenarios. Solid: fluctuation of the electron
injection power; dotted: fluctuation of electron spectral index $q$; 
long-dashed: fluctuation of $\gamma_2$; dot-dashed: fluctuation of both 
$q$ and $\gamma_2$ simultaneously. }
\label{sv_comparison}
\end{figure}

\newpage

$\to$ thick long-dashed;

\begin{figure}
\plotone{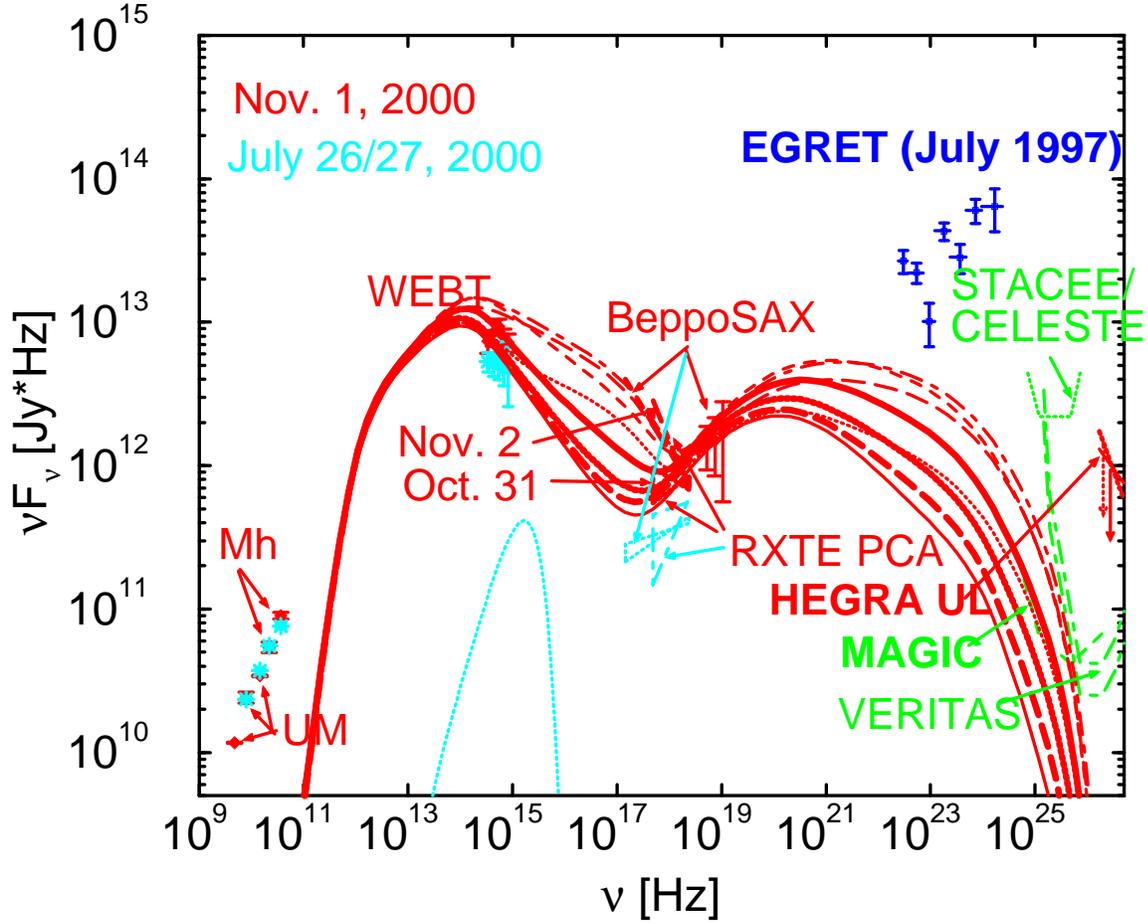}
\caption{Time-dependent model spectra for our combined SED + spectral
variability fit. Parameters: $D = 18$, $L_{\rm jet} = 2.5 \times
10^{40}$~ergs~s$^{-1}$, $\gamma_1 = 1000$, $\gamma_2 = 5 \times 10^4$,
$q = 3 \to 2.40$ from quiescent to flaring state. Time sequence is: 
thin solid $\to$ thin dotted $\to$ thin long-dashed $\to$ thin 
dot-dashed $\to$ thin dashed $\to$ thick solid $\to$ thick dotted
$\to$ thick long-dashed; equi-distant time steps of $\Delta t_{\rm obs} 
= 0.97$~hr.}
\label{mw_fit}
\end{figure}

\newpage

\begin{figure}
\plotone{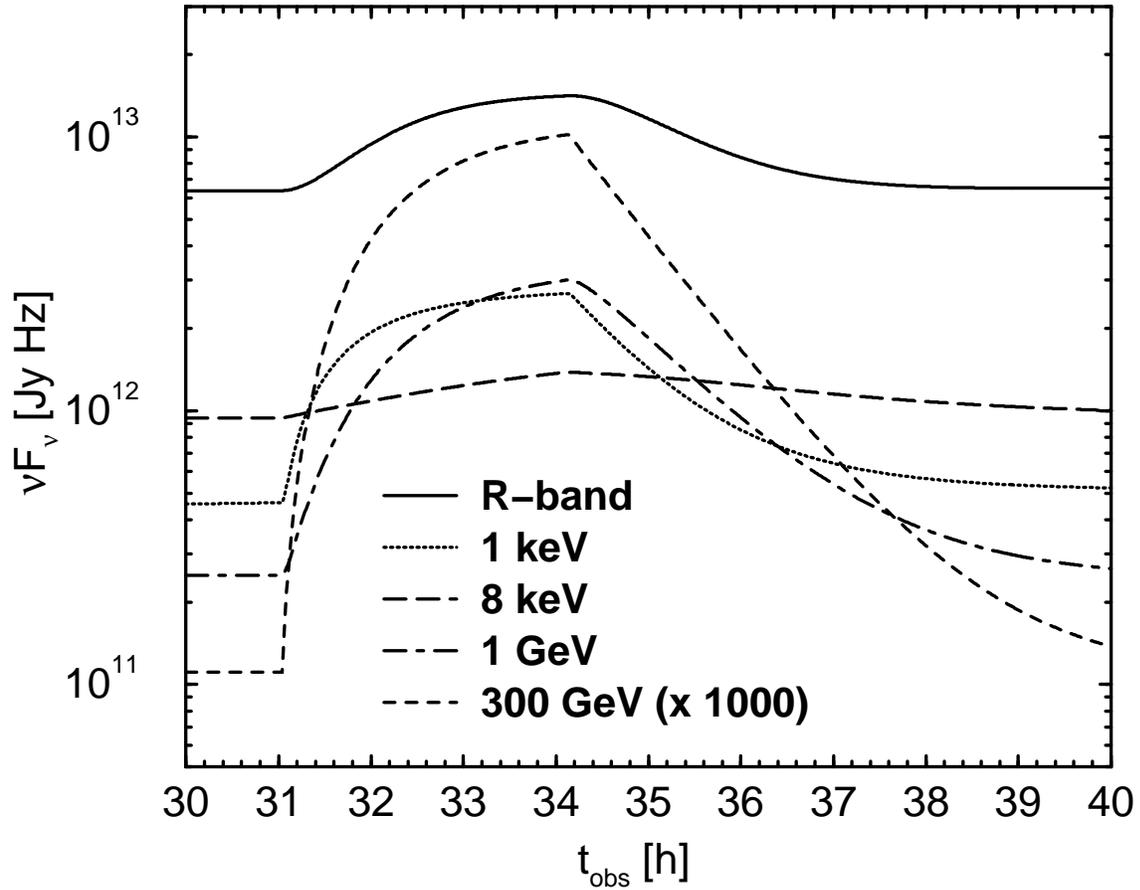}
\caption{Model light curves from the fit illustrated in Fig.~\ref{mw_fit}.
The 300~GeV light curve has been shifted up by a factor of $10^3$ in order
to fit on the same scale as the other light curves.}
\label{lightcurve_fit}
\end{figure}

\newpage

\begin{figure}
\plotone{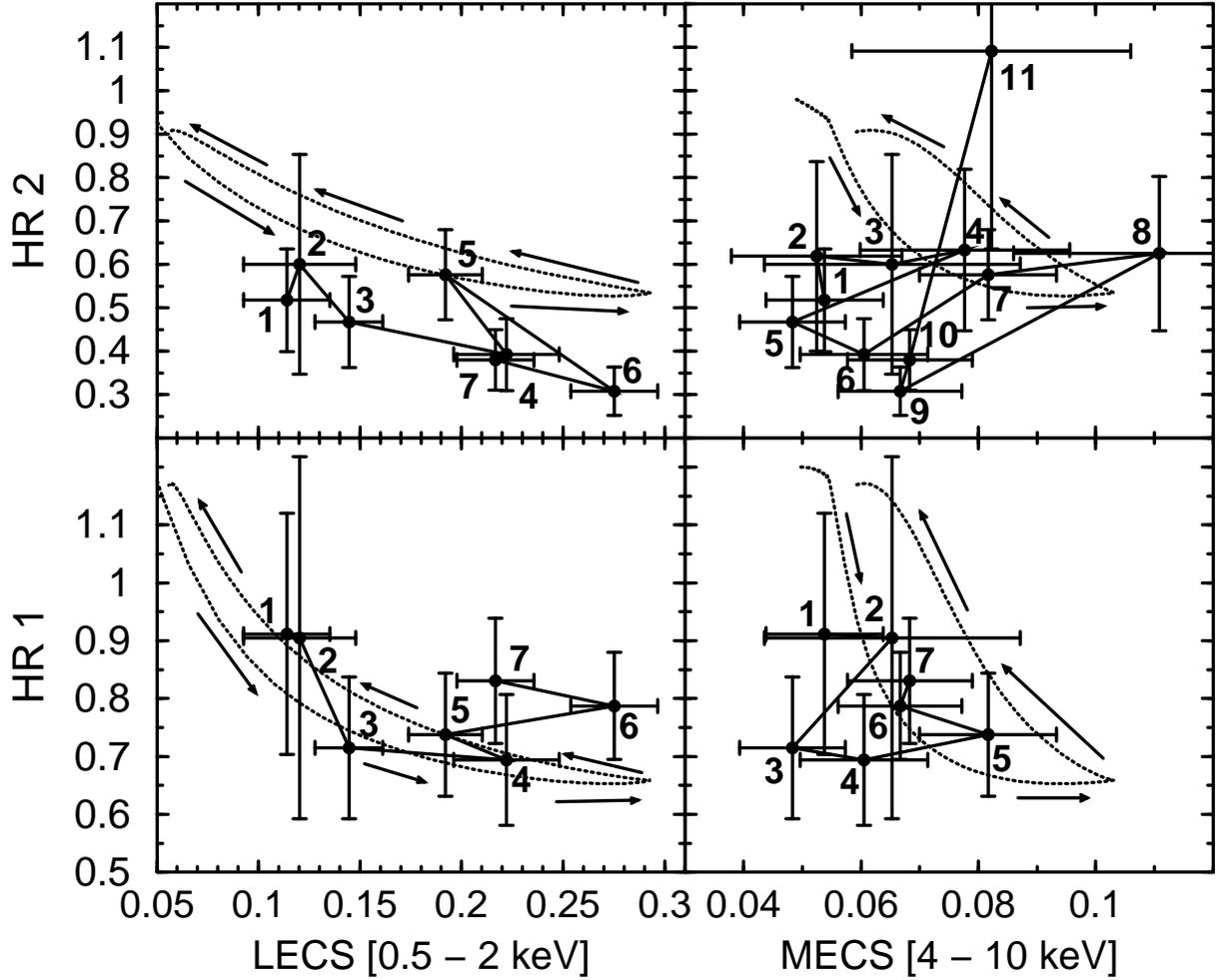}
\caption{Hardness-intensity diagram of the {\it BeppoSAX} hardness ratios
HR1 and HR2 as defined in \S \ref{observations} vs. soft X-ray LECS and 
medium-energy MECS flux for the well-resolved X-ray flare at t = 0.5 -- 
6.5~h of Nov. 1, 2000 \citep[data from][]{ravasio03}. The dotted curves 
indicate the simulated spectral hysteresis curves from our best-fit flaring 
scenario (Fig.~\ref{mw_fit}).}
\label{hid30_fit}
\end{figure}

\newpage

\begin{deluxetable}{cccccccccc}
\tabletypesize{\scriptsize}
\tablecaption{Fit parameters for the spectral fitting (equilibrium states)
to the SEDs of BL~Lacertae on July 26/27 and Nov. 1, 2000. The ``jet luminosity'' 
for leptonic models quoted below is the luminosity injected into relativistic 
electrons in the blob. If the quoted magnetic field is assumed to be present
throughout the jet (not only in the ``blob'' of relativistic electrons), the 
total jet luminosity will ultimately be dominated by the magnetic field energy 
density, $L_{\rm jet}^B \approx 6.1 \times 10^{42}$~ergs~s$^{-1}$. The jet
luminosities of the hadronic models are calculated following \cite{pm00}
which includes the magnetic field energy density.}
\tablewidth{0pt}
\tablehead{
\colhead{Model}         & \colhead{$\gamma_1$} & 
\colhead{$\gamma_2$}    & \colhead{$\gamma_{\rm p, max}$} & 
\colhead{$n_e/n_p$}     & \colhead{$q$}        & 
\colhead{$L_{\rm jet}$} & \colhead{$B$ [G]}    & 
\colhead{$R_{\rm B}$}   & \colhead{D} \\
\colhead{} & \colhead{(el.)} & \colhead{(el.)} & \colhead{} & \colhead{} & 
\colhead{} & \colhead{[ergs~s$^{-1}$]} & \colhead{} & \colhead{[cm]} & 
\colhead{} 
}
\startdata
Lept., July 26/27 & 1100 & $2.3 \times 10^4$ & ---             & --- & 2.4  & 
   $3 \times 10^{40}$ & 1.4 & $2.5 \times 10^{15}$ & 16 \\
Lept., Nov. 1     & 1100 & $6 \times 10^4$   & ---             & --- & 2.15 & 
   $4 \times 10^{40}$ & 1.4 & $2.5 \times 10^{15}$ & 18 \\
\noalign{\hrule} \\
Hadr., July 26/27 & 1 & $2.4\times 10^3$ & $9 \times 10^9$ & 2.7 & 1.9  &
   $6\times 10^{44}$ & 40  & $1.1 \times 10^{15}$ & 7  \\
Hadr., Nov. 1     & 1 & $2.1\times 10^4$ & $1.5\times 10^{10}$ & 1.9 & 1.9  &
    $7\times 10^{44}$ & 40  & $1.5 \times 10^{15}$ & 9  \\
\enddata
\label{equilibrium_fit}
\end{deluxetable}

\begin{deluxetable}{cccc}
\tabletypesize{\scriptsize}
\tablecaption{Predicted multi-GeV -- TeV fluxes from the spectral fits to
the SEDs of BL~Lacertae on July 26/27 and Nov. 1, 2000.}
\tablewidth{0pt}
\tablehead{
\colhead{Model}                      & \colhead{$\Phi_{> 5 \, {\rm GeV}}$} & 
\colhead{$\Phi_{> 40 \, {\rm GeV}}$} & \colhead{$\Phi_{> 100 \, {\rm GeV}}$} \\
\colhead{} & 
\colhead{[photons~cm$^{-2}$~s$^{-1}$]} & 
\colhead{[photons~cm$^{-2}$~s$^{-1}$]} & 
\colhead{[photons~cm$^{-2}$~s$^{-1}$]}
}
\startdata
Lept., July 26/27 & $1.6 \times 10^{-9}$ & $2.0 \times 10^{-11}$ & $1.2 \times 10^{-12}$ \\
Lept., Nov. 1     & $2.3 \times 10^{-9}$ & $7.2 \times 10^{-11}$ & $8.6 \times 10^{-12}$ \\
\noalign{\hrule} \\
Hadr., July 26/27 &  $1.1 \times 10^{-9}$ & $1.4-1.7 \times 10^{-10}$ & $2.9-4.0 \times 10^{-11}$\\
Hadr., Nov. 1     &  $0.9 \times 10^{-9}$ & $2.0-2.2 \times 10^{-10}$ & $4.7-6.7 \times 10^{-11}$\\
\enddata
\label{VHE_fluxes}
\end{deluxetable}

\end{document}